\documentclass[a4paper,10pt]{article}
\usepackage{graphicx}

\begin{document}


\begin{titlepage}

\begin{center}
\Large{\bf Canonical solution of classical magnetic
models with long-range couplings}
\end{center}
\vspace{2cm}

\begin{center}
\large{
Alessandro Campa,\footnote{
\normalsize{Physics Laboratory, Istituto Superiore di Sanit\`a and
INFN Sezione di Roma1, Gruppo Collegato Sanit\`a, Viale Regina
Elena 299, 00161 Roma, Italy}}
Andrea Giansanti,\footnote{
\normalsize{Physics Department, Universit\`a di Roma ``La Sapienza'' and
INFM Unit\`a di Roma1, Piazzale Aldo Moro 2, 00185 Roma, Italy}}$^{,\S}$
and Daniele Moroni\footnote{
\normalsize{Department of Chemical Engineering, Universiteit van Amsterdam,
Nieuwe Achtergracht 166, 1018 WV Amsterdam, The Netherlands}}
}
\end{center}
\vspace{5cm}

Short title: Canonical solution of long-range magnetic models

\vspace{2cm}
$^{\S}$Corresponding author\\
Address: Physics Department, Universit\`a di Roma ``La Sapienza'',
Piazzale Aldo Moro 2, 00185 Roma, Italy;\\
telephone: +39-06-49913491;\\
fax: +39-06-4463158;\\
electronic address: andrea.giansanti@roma1.infn.it
\end{titlepage}

\pagestyle{empty}
\begin{center}
{\bf Abstract}
\end{center}
\vspace{0.5cm}
We study the canonical solution of a family of classical $n-vector$ spin
models on a generic $d$-dimensional lattice; the couplings between two
spins decay as the inverse of their distance raised to the power $\alpha$,
with $\alpha<d$. The control of the thermodynamic limit requires the
introduction of a rescaling factor in the potential energy, which makes
the model extensive but not additive. A detailed analysis of the
asymptotic spectral properties of the matrix of couplings was necessary
to justify the saddle point method applied to the integration of functions
depending on a diverging number of variables. The properties of a class of
functions related to the modified Bessel functions had to be investigated.
For given $n$, and for any $\alpha$, $d$ and lattice geometry, the solution is
equivalent to that of the $\alpha=0$ model, where the dimensionality $d$ and
the geometry of the lattice are irrelevant.

\vspace{1cm}
\newpage
\pagestyle{plain}
\addtocounter{page}{-1}


\section{Introduction}\label{intro}
Systems with long-range interactions do not have a well defined thermodynamic
limit,$^{(1)}$ and their equilibrium statistical mechanics is not inside
the framework of the more studied short-range systems. For a pair interaction
long-range means, in this framework, that the modulus of the potential energy
decays, at large
distance, not faster than the inverse of the distance raised to the power of
the spatial dimension. In fact, conditions can be given for the existence of
the thermodynamic limit,$^{(2,3)}$ one being
\emph{temperedness}, that for pair interactions is exactly the requirement that
the coupling is not long-range as just defined, and the other being
\emph{stability}, which requires that the global minimum of
the potential energy does not diverge to $-\infty$ more than linearly with the
number of degrees of freedom.$^{(4)}$ The physically relevant example of
continuous long-range systems violating temperedness is that of particles
interacting with gravitational forces; here there is in addition the problem
of the potential energy at short distance, that manifestly violates stability.
Both conditions are necessary for the system to be \emph{extensive} and
\emph{additive}.
Extensivity means that the specific thermodynamic potentials (i.e., the
thermodynamic potentials per particle), do not diverge in the thermodynamic
limit, while additivity means that the thermdynamic potentials of the whole
system are, in that limit, the sum of those of its component macroscopic
parts.$^{(5,6)}$ In particular, additivity is essential if one wants to derive
the canonical ensemble from the microcanonical ensemble.

Since the classic paper by Ising,$^{(7)}$ magnetic models on a lattice
are the most used to investigate the statistical physics of interacting
many-body systems, because their mathematical tratment, although still
difficult, is often more affordable than that of continuous systems.
Many generalizations of the Ising model have been considered, over time;
among them, also models with long-range interactions. In lattice systems the
problem of the behaviour of the potential energy at short distance is not
present, and no collapse, as in gravitational models, can exist; the
equality of all the coordinates, as in the case of maximum magnetization, is
not a pathological configuration. The lack of stability and temperedness can
be related only to a coupling between spins decaying too slowly with distance.
Thus, the study of long-range magnetic systems on lattices can give insights on
the statistical properties in general of long-range potentials, and is
therefore useful to investigate how to extend the framework of statistical
mechanics to systems that do not obey stability and temperedness.

In this paper we study the equilibrium statistical mechanics of classical
$n-vector$ spins (i.e., $n$-dimensional unit vectors) fixed on the nodes of
a generic $d$-dimensional lattice, and interacting pairwise through a
long-range potential. We consider generic values of $n$ and $d$.
The couplings that we consider decay as the inverse of the distances between
spins raised to the power $\alpha$; if $\alpha$ is not larger than $d$ the
interaction is not tempered, and the specific energy (or energy density) of
the system diverges in the thermodynamic limit. This divergence can be cured
with a \emph{Kac's prescription}, that is gauging the potential
energy with an appropriate scaling function of $N$, the number of spins, and
$d$.$^{(8)}$ Then extensivity is enforced through a control of the
thermodynamic limit but, due to the long-range couplings, additivity does not
hold, and ensemble equivalence, whose proof is based on the possibility of
separating the energy of a subsystem from that of the whole, might not be
guaranteed in that limit.$^{(6)}$
We study our system in the canonical ensemble and we find the exact
soultion, and in subsection \ref{micro} we will show that for this class
of systems the microcanonical and the canonical ensembles are equivalent, in
spite of the non additivity.

The main motivation of this work is to show that all the systems included
in our study, and obtained for the different values of $n$, $d$ and for the
different possible geometries of the lattice, are essentially all equivalent
as far as their equilibrium statistical thermodynamics is concerned; some
quantitative differences (e.g., the value of the critical temperature) are
found between systems with different $n$, but the overall behaviour is always
the same. This is true also with regard to the dependence on the distance of
the coupling between spins, as long as it remains long-range: although we
consider a power law function, we argue in the final discussion that any other
form of the long-range coupling would have brought us to the same results.

The paper is divided into five sections. In Section 2 we present
the class of long-range magnetic models that we consider.
The canonical solution for the mean field case ($\alpha=0$) is given
in Section 3. Section 4 is devoted to the general mean field case
($0\leq \alpha <d$); it contains the main results 
of this paper and is divided into five subsections: in the first three the
canonical solution is obtained, while the last two are devoted to the
justification of the saddle point method and to the proof of the ensemble
equivalence. In the discussion of Section 5 we argue about possible
generalizations of our results.

\section{The Model}\label{themod}
We consider here classical models which belong to the general class of
the $n-vector$ models;$^{(9)}$ in our previous work$^{(10)}$ we had
studied a kind of $XY$-model, which is a vector model with $n=2$; the case
$n=3$ would correspond to a classical Heisenberg model. Classical $n-vector$
spin models correspond to the ``infinite spin'' limit of magnetic quantum
systems;$^{(11)}$ these models are related also to the spherical model
of Berlin and Kac:$^{(12)}$ in the limit $n\rightarrow\infty$ the
$n-vector$ model reduces to the spherical model.$^{(13)}$ The
consideration of long-range couplings in magnetic models on a lattice was
started in the sixties of the last century, when the basic mathematical
techniques were established.$^{(14)}$

The hamiltonian we consider is:
\begin{equation}\label{model0}
H=\frac{1}{2}\sum_{i,j=1}^N J_{ij} (1-\mathbf{S}_i \cdot \mathbf{S}_j).
\end{equation}
For each $i$ the spin $\mathbf{S}_i$ is a unit $n$-dimensional vector; its
position can be specified by $n-1$ angles and its cartesian coordinates are
related to these angles through the definition of the polar coordinates in
$\mathcal{R}^n$. The index $i=1,\ldots,N$ labels the sites of a generic
$d$-dimensional lattice, with $d$ integer. We conventionally put an external
factor $1/2$ and assume the rescaled form $1-\mathbf{S}_i \cdot \mathbf{S}_j$
which allows a free choice of the diagonal terms $J_{ii}$ because of the
constraint $\mathbf{S}_i^2=1$. We consider interactions in the form of an
inverse power of the distance $r_{ij}$ between lattice site $i$ and $j$:
\begin{equation}\label{Jij}
J_{ij}=\frac{1}{r_{ij}^\alpha} \qquad \qquad (i\neq j),
\end{equation}
with periodic boundary conditions and the nearest image convention for the
distance $r_{ij}$; $\alpha\geq 0$ sets the range of the interaction, which is
long-range if $\alpha \le d$ and short-range if $\alpha > d$.$^{(15)}$
Different models remain characterized by different values of $d,n$ and
$\alpha$. As said above, most of the work has been done in the study of models
with $\alpha > d$; $\alpha=\infty$ is the limit of nearest neighbors
interactions. We focus our attention instead on the case $\alpha<d$; in this
case the interaction (\ref{Jij}) causes the energy (\ref{model0}) to be not
extensive; thus the partition function does not admit a well defined
thermodynamic limit. The problem can be solved through the rescaling:
\begin{equation}\label{rescale}
J_{ij} \longrightarrow \frac{J_{ij}}{\tilde{N},}
\end{equation}
where $\tilde{N}$ is a function of $N$, $\alpha$, $d$ and the geometry of the
lattice. The above substitution is a generalization of the usual one found
in the mean field version of this kind of models: the general interaction
$J_{ij}$ is replaced by a constant one (i.e., $\alpha=0$), but a rescaling
factor $1/N$ is necessary to control the thermodynamic limit. The mean field
substitution is recovered as a particular case of (\ref{rescale}) for
$\alpha=0$. In the following we will sometimes refer to the general model,
with $0\le \alpha <d$, as the general mean field case.

We previously studied this rescaled model in the case $n=2$;$^{(10)}$ in that
short paper, where only few mathematical arguments and details were given, we
therefore considered only the model represented by planar rotators
on a lattice, but with the addition of a kinetic term in canonical variables,
conjugate to the angles. Here we will not consider the kinetic energy, that in
the canonical ensemble gives an additive trivial contribution to the
thermodynamic potentials; we will make very short comments on this in the
right places. Computation of the partition function for $n=2$ has shown
its universality in $\alpha$: the free energy does not depend on the value of
$\alpha<d$ and is thus equal to the mean field one for $\alpha=0$. The model
of planar
rotators, $n=2$, has also been the subject of some studies on the dynamical
properties: the universality of the thermodynamics in a one-dimensional
lattice ($d=1$) with respect to $\alpha<1$ was suggested by the numerical
study in ref. 16; interesting metastable states have been observed for
$\alpha=0$$^{(17,18)}$ and for $\alpha<1$ with $d=1$.$^{(19)}$

Here we take a general $n-vector$ model and we show how to treat the
interaction $1/r_{ij}^\alpha$ in the case $\alpha<d$ as far as the
computation of the partition function is concerned.

\section{The mean field case $\alpha=0$}\label{MeanField}

In this Section we review the canonical solution of the $\alpha=0$ model; in
the case of planar rotators, that is $n=2$, the solution has already been
given in ref. 20. For $\alpha=0$ the interaction is homogeneous, i.e.,
each spin interacts with all the others with the same strength.
In this case the rescaling (\ref{rescale}) is usually performed with
$\tilde{N}=N$. The hamiltonian is then:
\begin{equation}\label{MF.H}
H=\frac{1}{2N}\sum_{i,j=1}^N (1-\mathbf{S}_i \cdot
\mathbf{S}_j) - \mathbf{h} \cdot \sum_{i=1}^N \mathbf{S}_i,
\end{equation}
where we also include an external magnetic field $\mathbf{h}$, which is
coupled in the usual way to the spin vectors $\mathbf{S}_i$.
The partition function for the model is
\begin{equation}\label{MF.Z0}
Z=\int d^N\Theta \: e^{-\beta H},
\end{equation}
where $\beta$ is the inverse temperature, $d\Theta_i$ is the surface element
of the unit sphere in dimension $n\geq 2$, and $d^N\Theta=d\Theta_1\ldots
d\Theta_N$; and for $n=1$ the integral is replaced by a sum on all the
possible Ising spin configurations: $\sum_{S_1=\pm 1,\ldots,S_N=\pm 1}
e^{-\beta H}$. Here, and later for the general case, we do not consider the
kinetic part, that in the classical partition function trivially decouples.
The magnetization is given by:
\begin{equation}\label{magnet}
\mathbf{M}\equiv\frac{1}{N}\langle\sum_{i=1}^N
\mathbf{S}_i\rangle=\frac{1}{N\beta}\frac{\partial}{\partial
\mathbf{h}}\ln Z,
\end{equation}
where $\langle \cdot \rangle =\frac{1}{Z}\int d^N\Theta (\cdot) e^{-\beta H}$
is the usual canonical average. With $\mathbf{B}=\beta \mathbf{h}$ and
$C=\exp(-N\beta/2)$, we can rewrite:
\begin{equation}\label{MF.Z1}
Z=C \int d^N\Theta \: e^{\frac{\beta}{2N}\left|\sum_i \mathbf{S}_i\right|^2
+\mathbf{B}\cdot\sum_i\mathbf{S}_i}.
\end{equation}
Using the gaussian transformation
\begin{equation}\label{gauss}
 e^{a S^2} = \frac{1}{\sqrt{4 \pi a}} \int_{- \infty}^{+ \infty} dz
 e^{-\frac{z^2}{4a} + S z} \qquad a>0
\end{equation}
on each term of the square modulus of the vector $\sum_i\mathbf{S}_i$
(we emphasize that the above expression is valid when $a>0$, or more generally,
for complex $a$, when its real part is greater than $0$), we linearize the
quadratic term in (\ref{MF.Z1}) and obtain:
\begin{equation}\label{MF.Z2}
Z=C \left(\frac{N}{2\pi\beta}\right)^{n/2} \int d \mathbf{z} \:
e^{-\frac{N}{2\beta}\mathbf{z}^2}
\int d^N\Theta \: e^{\sum_i \mathbf{S}_i \cdot (\mathbf{z}+\mathbf{B})},
\end{equation}
where $d\mathbf{z}=dz_1\dots dz_n$. Here and in the following the notation
$\mathbf{b}^2$ will denote, interchangeably with $b^2$, the scalar product of
the vector $\mathbf{b}$ with
itself, i.e., its square modulus. The last integral separates on the sites
$i$ and gives $N$ identical
contributions, the functional form of which depends on the spin dimension $n$.
In Appendix A we show a more convenient way to write these integrals on the
unit sphere and how they are related to the modified Bessel functions; besides,
we introduce the notation, that expresses the surface integrals in
(\ref{MF.Z2}) in terms (for proper values of $n$) of a function $G_n(x)$ and of
the area $\Omega_n$ of the unit sphere in $n$ dimensions. Some properties of
the functions defined in Appendix A are needed for the analysis of this mean
field case and of the more general case with $\alpha \neq 0$; these properties
are proved in Appendix B. Following the notation introduced above
we rewrite the partition function as:
\begin{equation}\label{MF.Z3}
Z=C \Omega_{n-1}^N \left(\frac{N}{2\pi\beta}\right)^{n/2} \int d \mathbf{z}
 \: e^{N\left[-\frac{1}{2\beta}\mathbf{z}^2 + \ln G_{n-2}\left(
|\mathbf{z}+\mathbf{B}|\right)\right]}.
\end{equation}
The integral is computed with the saddle point method; therefore, we need to
find the stationary points of the function in square brackets in the
exponent, and consider those that are maxima; the dominant contribution
to the integral will be determined by the absolute maximum. 
If we call $f(\mathbf{z})$ the function in square brackets in (\ref{MF.Z3}),
the stationary points are given by the solutions of the system of $n$
equations (one for each component $z_\mu$ of $\mathbf{z}$):
\begin{equation}\label{mfsol1t}
\frac{\partial f}{\partial z_\mu}=-\frac{1}{\beta}z_\mu + g_{n-2}
(|\mathbf{z}+\mathbf{B}|)\frac{z_\mu+B_\mu}{|\mathbf{z}+\mathbf{B}|}=0,
\quad \quad \mu=1,2,\dots,n
\end{equation}
where the function $g$ is the logarithmic derivative of $G$. The free energy
per particle (or specific free energy) will be given by:
\begin{eqnarray}\label{freemf}
-\beta F &=&\lim_{N\rightarrow\infty}\frac{1}{N}\ln Z= \nonumber \\
&=&-\frac{\beta}{2}+
\ln \Omega_{n-1}+\max_{\mathbf{z}}\left[-\frac{1}{2\beta}\mathbf{z}^2 +
\ln G_{n-2}\left(|\mathbf{z}+\mathbf{B}|\right)\right] \nonumber \\
&\equiv&-\frac{\beta}{2}+
\ln \Omega_{n-1}+\max_{\mathbf{z}}f(\mathbf{z}).
\end{eqnarray}
We note that the hessian of $f(\mathbf{z})$ in the maximum does not appear in
(\ref{freemf}), since its contribution becomes vanishingly small in the
thermodymamic limit, $N\rightarrow \infty$. An analogous
argument can be used for possible degeneracies of the absolute maximum (see
few lines below). The study of
(\ref{mfsol1t}) is presented in some details in Appendix C; here we only show
the results and the final expressions for the magnetization and for the
equation of state. Eq. (\ref{mfsol1t}) can have more than one solution,
depending on the value of $\beta$ and $B=|\mathbf{B}|$. In any case, the
relevant stationary point $\mathbf{z}^*$ is such that its modulus $z^*$
satisfies the self-consistency equation:
\begin{equation}\label{SC}
z^*=\beta g_{n-2}(z^*+B),
\end{equation}
which is a generalization of the Curie-Weiss equation found in the
mean-field solution of the Ising model. In fact, as shown in Appendix
B, $g_n(x)$ has the same qualitative features of $\tanh (x)$.
A visual aid (Fig. 1) is provided in Appendix C together with the details
of the study of Eq. (\ref{mfsol1t}).
To complete the solution, we must add the following specifications. When $B>0$
(i.e., $h=|\mathbf{h}|>0$), for which (\ref{SC}) has a positive solution
$z^*>0$, $\mathbf{z}^*$ is parallel to
$\mathbf{h}$. When $h=0$, we have to distinguish between $\beta > \beta_c=n$
and $\beta \leq \beta_c$: for inverse temperatures not greater than the
critical value $\beta_c$ the only solution of (\ref{SC}) for $h=0$ is $z^*=0$;
instead, for $\beta>\beta_c$ there is also a positive solution, and this is the
relevant one. In this last case, the direction of $\mathbf{z}^*$ (if $n>1$)
remains undetermined; in other words, the stationary point is infinitely
degenerate, or doubly degenerate if $n=1$. Therefore, to be more precise, one
should then perform in (\ref{MF.Z3}), when $\beta>\beta_c$ and $h=0$, an
integration over the angular coordinates of $\mathbf{z}$ (or a sum over the two
directions if $n=1$) before applying the saddle point. This would give in
(\ref{freemf}) a further factor $\frac{\ln \Omega_n}{N}$, which does not
contribute to $F$ in the thermodynamic limit.

The magnetization $\mathbf{M}$ is given by (\ref{magnet}), that applied to
(\ref{freemf}) gives $\mathbf{M}=\frac{\mathbf{z}^*}{\beta}$. Since in the case
$h=0$ and $\beta>\beta_c$ the direction of $\mathbf{z}^*$ is not determined,
because of degeneracy, the actual direction of $\mathbf{M}$ in a real
system is determined by the boundary conditions, and there is a spontaneous
simmetry breaking. The magnetization modulus $M=|\mathbf{M}|$ becomes zero at
$\beta=\beta_c$ and remains zero for $\beta<\beta_c$; at $\beta_c$ there is a
second order phase transition.

Finally, the equation of state, relating the specific (potential) energy
$U=\frac{1}{N}\langle H \rangle$ to the temperature (through the
magnetization modulus $M$), is given by:
\begin{equation}\label{eqstmf}
U=-\lim_{N\rightarrow\infty}\frac{1}{N}\frac{\partial}
{\partial \beta}\ln Z=\frac{1}{2}\left(1-M^2\right)-hM.
\end{equation}
If we had considered also the kinetic energy, then in (\ref{freemf}) we would
have had an additional term $\frac{n-1}{2}\ln\left(\frac{2\pi}{\beta}\right)$,
and consequently in (\ref{eqstmf}) a further term $\frac{n-1}{2\beta}$ would
have appeared, making  in that case $U$ the total specific energy. For
$\beta \le \beta_c$ the specific potential energy remains constant equal to
$\frac{1}{2}$ (since $M$ remains equal to zero), and only the specific kinetic
energy increases.

The phase space volume at disposition of the rotators increases with $n$, and
this is reflected in a critical temperature $T_c=\frac{1}{n}$ decreasing with
$n$; apart from this quantitative difference, the overall behaviour is the same
for all values of $n$.

\section{The general mean field case $0\leq \alpha <d$}\label{Interaction}
This is the central Section of the paper. We take into consideration the
general form of the interaction, i.e. $J_{ij}=1/r_{ij}^\alpha$ with
$0\le \alpha <d$ (see (\ref{model0}) and (\ref{Jij})). From the form of $H$ it
is clear, since $\mathbf{S}_i$ is a unit vector, that the values of $J_{ii}$
can be chosen arbitrarily, as long as they are finite. We will use this
freedom below, for the computation of the partition function.

We do not explicitly consider the marginal case $\alpha=d$.
However, in the discussion we will comment on this point and on possible more
general forms of the long-range couplings.

\subsection{Thermodynamic limit}\label{therlim}
We have to consider the problem of the rescaling of the interaction
parameters, to control the thermodynamic limit. For the classical lattice
systems considered here the existence of the thermodynamic limit
is guaranteed by a restriction on the long-range part of the potential energy.
For translationally invariant interactions, i.e., when in our case $J_{ij}$
depends only on the distance vector from site $i$ to site $j$, the restriction
takes the form:
\begin{equation}\label{TL}
\lim_{N\rightarrow\infty} \sum_{j=1}^N |J_{ij}| < \infty,
\end{equation}
where $i$ can take any value, since the translational invariance implies that
the above sum is independent of $i$. It corresponds to an
extensivity requirement for the energy $H$. The rigorous demonstration of the
sufficiency of this condition for the existence of the thermodynamic limit of
the partition function can be found in ref. 2, where the Ising model
($n$=1) is considered, but the procedure can be extended to the general case.

For our system we introduce translational invariance also for finite $N$,
through the use of periodic boundary conditions and a choice of the same finite
value $b$ for all the diagonal terms $J_{ii}$; this is convenient for many
steps of the analysis of our model. Apart from this,
condition (\ref{TL}) requires the analysis of the quantity
\begin{equation}\label{range}
S=\sum_{j\neq i}\frac{1}{r_{ij}^\alpha}
\end{equation}
in the $N\rightarrow\infty$ limit. We have dropped the $j=i$
term, since the single value of $J_{ii}$ does not determine the convergence
property of (\ref{range}). To see the large $N$ behavior of (\ref{range})
for $\alpha<d$ we can shift to the integral, that gives $S\sim N^{1-\alpha/d}$
(see, e.g., next subsection). Divergence occurs and must be
compensated to have a thermodynamically well defined model. Compensation is
readily obtained with the position (\ref{rescale}), taking
\begin{equation}\label{ntilde0}
\tilde{N}\sim S.
\end{equation}

\subsection{Interaction matrix (spectral properties)}\label{intermat}
The nonrescaled couplings $J_{ij}$ are the entries of a real, symmetric
$N\times N$ matrix. In this subsection we study its eigenvalues, since this
is important for
the computation of the partition function. This matrix can be diagonalized
through a unitary matrix $V$ and its eigenvalues are all real. In our
computation of the partition function it will be important that all the
eigenvalues are positive, since we will make use of the gaussian transformation
(\ref{gauss}). We show here how we can use the freedom on the value
of $J_{ii}$ for this purpose.

Let us denote the position of lattice site $i$ by $\mathbf{r}_i$. We stress
that in this subsection, and only in this one, boldface characters denote
$d$-dimensional vectors of the lattice space or of its dual; in the rest of the
paper, before and after this subsection, they denote $n$-dimensional vectors
related to the dimensionality of the spins. The
function $r_{ij}$ is the distance between the point $\mathbf{r}_i$ and the
nearest image of $\mathbf{r}_j$, and then it is invariant under
translations. The same is true for $J_{ij}$, once
we use the freedom on the values of $J_{ii}$ taking all these diagonal
elements equal to the same constant $b$. Therefore, if we let
$\mathbf{r}_{ij}=\mathbf{r}_i-\mathbf{r}^{(i)}_j$, where $\mathbf{r}^{(i)}_j$
is the image of $\mathbf{r}_i$ which is nearest to $\mathbf{r}_j$, denoting
$r_{ij}=|\mathbf{r}_{ij}|$ and introducing the notation $J_{ij}=J(r_{ij})$,
we have:
\begin{equation}
J(r_{ij})=\left\{
\begin{array}{rl}
b & \mbox{if } r_{ij}=0 \\
r_{ij}^{-\alpha} & \mbox{ otherwise.}
\end{array}
\right.
\end{equation}
Following these definitions, the eigenvalues of the matrix $J$
are obtained through the $d$-dimensional Fourier transform
\begin{equation}\label{lambdas}
\lambda_\mathbf{k}=\sum_\mathbf{r} J(r) e^{-i \mathbf{k} \cdot \mathbf{r}}=
\sum_{\mathbf{r}} r^{-\alpha} e^{-i \mathbf{k} \cdot \mathbf{r}},
\end{equation}
where the sum is on all the lattice points and $\mathbf{k}$ denotes any of the
$N$ reciprocal lattice vectors contained in the first Brillouin zone. The
reality of the eigenvalues follows from (\ref{lambdas}) and the properties of
$J(r_{ij})$. It is also evident that $\lambda_{\mathbf{0}}$ is the largest
eigenvalue. If we isolate the $r=0$ term we have:
\begin{equation}\label{translation}
\lambda_\mathbf{k}=b+\sum_{\mathbf{r}\neq \mathbf{0}} r^{-\alpha}
e^{-i \mathbf{k} \cdot \mathbf{r}},
\end{equation}
that shows that the whole spectrum is linearly translated by $b$. The
remaining sum for $\mathbf{k}=\mathbf{0}$ corresponds to the sum $S$ defined
in (\ref{range}). It is clear that for $\alpha>d$ all the eigenvalues are
finite in the thermodynamic limit. We now restrict to the case $\alpha<d$.
The large $N$ behaviour of $S$ can be estimated shifting $S$ to an integral
\begin{equation}\label{lambdaest1}
S=\sum_{r\neq 0} r^{-\alpha}\sim\int_1^{N^{\frac{1}{d}}} r^{d-1-\alpha} dr
\sim N^{1-\frac{\alpha}{d}}.
\end{equation}
If we take
\begin{equation}\label{ntilde1}
\tilde{N}=\lambda_{\mathbf{0}} = b+S = \sum_{j=1}^N J_{ij},
\end{equation}
the requirement (\ref{ntilde0}) is satisfied as an equality in the
thermodynamic limit because $b$ is a finite quantity; as noted before, the last
expression in (\ref{ntilde1}) does not depend on $i$. It is also possible to
estimate the behavior of $\lambda_\mathbf{k}$ for $\mathbf{k} \neq
\mathbf{0}$ for large $N$. It is easy to see that the sum in
(\ref{translation}), if $k=|\mathbf{k}|$ is different from $0$, remains finite
in the thermodynamic limit, and the behavior in $k$, again shifting to an
integral, can be found to be
\begin{equation}\label{lambdaest2}
\sum_{\mathbf{r}\neq \mathbf{0}} r^{-\alpha} e^{-i \mathbf{k} \cdot
\mathbf{r}}\sim \frac{1}{k^{d-\alpha}} \, .
\end{equation}
This expression does not consider the sign of the left hand side. The maximum
value of $k$ is of the order of the inverse of the lattice spacing, and in
the thermodynamic limit the $N$ vectors $\mathbf{k}$ tend to fill uniformly
a d-dimensional sphere with a radius equal to this maximum value. Therefore,
in this limit the possible values of $k$ are distributed according to
$k^{d-1}$. It follows that in the thermodynamic limit only a vanishing
fraction of the eigenvalues diverges (and at most as $\tilde{N}$, like
$\lambda_{\mathbf{0}}$), also in the less favourable case (concerning the
distribution of the values of $k$ near $0$), when $d=1$.
This will be important for our computation of the
partition function. It is also evident that negative eigenvalues are possible
only for finite values of $k$, and that the least eigenvalue, in the
case $b=0$ in (\ref{lambdas}), is of order one in modulus. Thus, the all
spectrum can be made positive by properly choosing the value of $b$ in
(\ref{lambdas}).

We can now estimate the behaviour of the rescaled eigenvalues pertaining to
the interaction (\ref{rescale}). The eigenvalues of the rescaled interaction
matrix are related to the $\lambda_\mathbf{k}$, being given by
$\lambda_\mathbf{k}/\tilde{N}$. If we choose a value of $b$ such that the
least eigenvalue $\lambda_\mathbf{k}$ has a positive value $\epsilon$, then
for the eigenvalues of the rescaled interaction the following relation holds:
\begin{equation}\label{spectrumrel}
0<\frac{\epsilon}{\tilde{N}}\leq \frac{\lambda_\mathbf{k}}{\tilde{N}} \leq 1.
\end{equation}
According to what has been noted above, only a vanishing fraction will remain
finite in the thermodynamic limit; this will be an important fact in the
following.

To end this subsection, we note that when $\alpha=0$ the eigenvalues can be
calculated from (\ref{translation}):
\begin{equation}
\lambda_\mathbf{k} = b + \sum_{\mathbf{r} \neq \mathbf{0}} e^{-i \mathbf{k}
\cdot \mathbf{r}} = b - 1 + N \delta_{k0}.
\end{equation}
There are $N-1$ eigenvalues equal to $b-1$ and one eigenvalue equal to
$b+N-1=\tilde{N}$. All the rescaled eigenvalues vanish in the
thermodynamic limit except the single biggest one equal to 1. This is just
the extreme case of the previous analysis.

\subsection{Canonical solution}\label{Solution}

We now show that thermodynamic universality holds among all
the rescaled $\alpha<d$ cases, in the sense that the specific free energy and
the equation of state are the same.

With the help of the analysis of the previous subsection, we are now able to
find an exact solution of the rescaled model defined by the hamiltonian:
\begin{equation}\label{H}
H=\frac{1}{2\tilde{N}}\sum_{ij=1}^N \frac{1-\mathbf{S}_i \cdot
\mathbf{S}_j}{r_{ij}^\alpha} - \mathbf{h} \cdot \sum_{i=1}^N \mathbf{S}_i,
\end{equation}
with $\alpha<d$, $J_{ij}=1/r_{ij}^\alpha$, $\tilde{N}$ defined as in
(\ref{ntilde1}), and $1/r_{ii}^\alpha \equiv b$ defined so as to have
(\ref{spectrumrel}). As in the mean field case, we have added an external
magnetic field $\mathbf{h}$.

In the following, whenever a site-dependent $n$-dimensional vector like
$\mathbf{S}_i$ occurs we indicate its components with double index
quantities $S_{i\mu}$ with the roman index varying in the range $1\ldots N$
and the greek index in the range $1\ldots n$. The $n$-dimensional vector at
site $i$ is indicated with boldface letters $\mathbf{S}_i$ while we let
$\underline{S}^T_\mu$ indicate the $N$-dimensional row vector
$(S_{1\mu},\ldots,S_{N\mu})$, and $\underline{S}_\mu$ the corresponding
transposed column vector. The components of a site-independent $n$-dimensional
vector, like $\mathbf{h}$ in (\ref{H}), will be indicated with, e.g., $h_\mu$.

The partition function for the model is given by eq.~(\ref{MF.Z0}) with the
new hamiltonian. Analogously to the mean field case, we define
$B_{\mu}=\beta h_{\mu}$ and
$C=\exp{[-(\beta/2\tilde{N})\sum_{ij} J_{ij}]}=\exp(-N\beta/2)$; due to the
site independence of the magnetic field, we have $\underline{B}^T_\mu=
(B_\mu,\ldots,B_\mu)$.
Introducing the matrix $R_{ij}=(\beta/2\tilde{N}) J_{ij}$ we can rewrite the
partition function in matrix form:
\begin{equation}\label{Z1}
Z=C \int d^N\Theta \: e^{\sum_\mu(\underline{S}_\mu^T R \underline{S}_\mu +
\underline{S}_\mu^T \underline{B}_\mu)}.
\end{equation}
Like in the mean field case we want to make use of gaussian transformations
to linearize the quadratic term. Then we first diagonalize the matrix $R$
with the unitary matrix $V$ such that $VRV^T=D$, with $D_{ij}=R_i\delta_{ij}$,
where $R_i$, the eigenvalues of $R$, are related to the eigenvalues $\lambda_i$
of $J_{ij}$ by $R_i=(\beta/2\tilde{N})\lambda_i$. So we can write the first
part of the exponent as
\begin{equation}\label{diag}
\sum_\mu \underline{S}_\mu^T R \underline{S}_\mu = \sum_{i\mu} R_i
\sigma_{i\mu}^2,
\end{equation}
where $\underline{\sigma}_\mu=V \underline{S}_\mu$. Because of
(\ref{spectrumrel}) all the eigenvalues $R_i$ are positive and we can apply
the gaussian transformation (\ref{gauss}) to each term in the right-hand side
of (\ref{diag}), obtaining:
\begin{equation}\label{gausstr}
e^{\sum_{i\mu} R_i \sigma_{i\mu}^2}=\frac{1}{[(4\pi)^N
\det R]^{\frac{n}{2}}} \int \Big( \prod_{i\mu}d v_{i\mu}\Big) e^{\left[
-\frac{v_{i\mu}^2}{4R_i}+\sigma_{i\mu}v_{i\mu}\right]} \, ,
\end{equation}
where the appearance of $\det R$ in the denominator is due to the relation
$\prod_i R_i = \det R$. Noting that $R_i^{-1}=(D^{-1})_{ii}$, with the change
of variables defined by $\underline{v}_\mu=V \underline{z}_\mu$, and
introducing the notation $\prod_{i\mu} dz_{i\mu}=d^{Nn} z$, the previous
expression can be written as:
\begin{equation}\label{gausstrb}
e^{\sum_{i\mu} R_i \sigma_{i\mu}^2}=\frac{1}{[(4\pi)^N
\det R]^{\frac{n}{2}}} \int d^{Nn} z e^{\left[-\frac{1}{4}\sum_\mu 
\underline{z}_\mu^T R^{-1}\underline{z}_\mu +\sum_\mu \underline{S}_\mu^T
\underline{z}_\mu\right]} \, .
\end{equation}
Inserting this in (\ref{Z1}), and noting that
$\sum_\mu \underline{S}_\mu^T (\underline{z}_\mu+\underline{B}_\mu)=
\sum_i \mathbf{S}_i\cdot (\mathbf{z}_i+\mathbf{B})$, the partition function
becomes:
\begin{equation}\label{Z1b}
Z=\frac{C}{[(4\pi)^N\det R]^{\frac{n}{2}}}
 \int d^{Nn} z e^{-\frac{1}{4}\sum_\mu \underline{z}_\mu^T R^{-1}
 \underline{z}_\mu}
\int d^N\Theta \: e^{\sum_i \mathbf{S}_i\cdot (\mathbf{z}_i+\mathbf{B})} \, .
\end{equation}
This expression is similar to (\ref{MF.Z2}), but with an n-component
integration variable $\mathbf{z}$ for each site $i$. With the same
notations used in Section \ref{MeanField} and discussed in the Appendices,
we therefore obtain:
\begin{equation}\label{Z1c}
Z=\frac{C\Omega_{n-1}^N}{[(4\pi)^N\det R]^{\frac{n}{2}}}
\int d^{Nn} z e^{\left[-\frac{1}{4}\sum_\mu \underline{z}_\mu^T R^{-1}
\underline{z}_\mu
+\sum_i \ln G_{n-2}\left(|\mathbf{z}_i+\mathbf{B}|\right)\right]} \, .
\end{equation}
This integral, analogous to (\ref{MF.Z3}) of the mean field case, can be
solved with the saddle point method. However, here the justification requires
more attention than before, since together with $N$ also the number of
integration variables becomes very large. We postpone this point to the next
subsection, and for the moment we look for the stationary points of the
exponent of (\ref{Z1c}) and we perform the stability analysis. We do not put
in evidence a factor $N$, since this is not necessary for the search of
the stationary points and for the determination of their character. If we
call $f(\{z_{i\mu}\})$ the exponent, the stationary points
are given by the solutions of the system of $Nn$ equations:
\begin{eqnarray}\label{solst1}
\frac{\partial f}{\partial z_{i\mu}}&=&-\frac{1}{2}\sum_{j=1}^N (R^{-1})_{ij}
z_{j\mu} + g_{n-2}(|\mathbf{z}_i+\mathbf{B}|)
\frac{z_{i\mu}+B_\mu}{|\mathbf{z}_i+\mathbf{B}|}=0 \nonumber \\
&&\mu=1,\dots,n \quad \quad i=1,\ldots,N.
\end{eqnarray}
The value of the integral in (\ref{Z1c}) will be determined by the absolute
maximum of $f(\{z_{i\mu}\})$; therefore, in the thermodynamic limit we will
obtain for the specific free energy:
\begin{eqnarray}\label{free1}
-\beta F&=&\lim_{N\rightarrow \infty}\frac{1}{N}\ln Z = -\frac{\beta}{2}
+\ln \Omega_{n-1} -n\ln 2 \\ &+&\lim_{N\rightarrow \infty}\frac{1}{N}
\left[ \max_{\{z_{i\mu}\}}f -\frac{n}{2}\ln \det R -\frac{1}{2} \ln
\det \left(-\frac{1}{2}H_0\right)\right], \nonumber
\end{eqnarray}
where $H_0$ is the hessian of $f$ computed in the absolute maximum (see below
how it has to be interpreted in case of degeneracy of this maximum). The
essential difference with respect to the expression of the partition function
of the $\alpha=0$ case, Eq. (\ref{freemf}), is represented by the terms
containing $\det H_0$ and $\det R$. There it was not
necessary to consider the hessian of the exponent; the reason is that
(\ref{MF.Z3}) is $n$-dimensional, and the contribution of the hessian to the
specific free energy vanishes in the thermodynamic limit. The second
derivatives of $f(\{z_{i\mu}\})$, necessary to compute the hessian, are given
by:
\begin{eqnarray}\label{secder}
&&\frac{\partial^2 f}{\partial z_{i\mu} \partial z_{j\nu}}= \nonumber \\
&=-&\frac{1}{2}
(R^{-1})_{ij}\delta_{\mu\nu} + \delta_{ij}\left[ g'_{n-2}(|\mathbf{z}_i+
\mathbf{B}|)\frac{(z_{i\mu}+B_\mu)(z_{i\nu}+B_\nu)}
{|\mathbf{z}_i+\mathbf{B}|^2} \right. + \\
&-& \left. g_{n-2}
(|\mathbf{z}_i+\mathbf{B}|) \frac{(z_{i\mu}+B_\mu)(z_{i\nu}+B_\nu)}
{|\mathbf{z}_i+\mathbf{B}|^3} + g_{n-2}(|\mathbf{z}_i+\mathbf{B}|)
\frac{\delta_{\mu\nu}}{|\mathbf{z}_i+\mathbf{B}|}\right]. \nonumber
\end{eqnarray}
The system (\ref{solst1}) is equivalent to the following:
\begin{eqnarray}\label{solst2}
z_{i\mu}&=&2 \sum_{j=1}^N R_{ij} g_{n-2}(|\mathbf{z}_j+\mathbf{B}|)
\frac{z_{j\mu}+B_\mu}{|\mathbf{z}_j+\mathbf{B}|} \nonumber \\
&&\mu=1,\dots,n \quad \quad i=1,\ldots,N.
\end{eqnarray}

\subsubsection{Homogeneous solutions}
Anticipating that the relevant stationary point is homogeneous on the
lattice, we show what is obtained if we look for a solution in which
$z_{i\mu}$ does not depend on $i$. Then (\ref{solst2}) reduces to the
system:
\begin{equation}\label{solst3}
z_\mu=\beta g_{n-2}(|\mathbf{z}+\mathbf{B}|)
\frac{z_\mu+B_\mu}{|\mathbf{z}+\mathbf{B}|} \quad \quad
\mu=1,\dots,n,
\end{equation}
which is identical to (\ref{mfsol1t}); in fact, from the definition of the
matrix $R_{ij}$ and from (\ref{ntilde1}) we have that $2\sum_j R_{ij}=\beta$
for each $i$. We take the solution of (\ref{solst3}) already considered for
$\alpha=0$, and we show that the stability analysis, now applied to the matrix
(\ref{secder}), gives the same results as for that case. Afterwards, we will
consider other possible solutions of (\ref{solst2}), non homogeneous on the
lattice, and we will show that other possible maxima, among them, are only
local.

In the case with magnetic field we considered, without loss of generality,
the vector $\mathbf{h}$ in the positive direction of the first axis, and we
found that the corresponding solutions of (\ref{solst3}) have $z_\mu=0$ for
$\mu\ne 1$. For such $\mathbf{z}$s, (and for $B_\mu=B\delta_{\mu 1}$),
the second derivatives vanish if $\mu \neq \nu$, while for the others
we have:
\begin{eqnarray}\label{secder1}
\left.\left(\frac{\partial^2 f}{\partial z_{i1} \partial z_{j1}}\right)
\right|_{z_{i\mu}=z_1\delta_{\mu 1}}&=& -\frac{1}{2}
(R^{-1})_{ij} + \delta_{ij}g'_{n-2}(|z_1+B|) \\
\left.\left(\frac{\partial^2 f}{\partial z_{i\mu} \partial z_{j\mu}}\right)
\right|_{z_{i\mu}=z_1\delta_{\mu 1}}&=&
-\frac{1}{2}(R^{-1})_{ij} +\delta_{ij}\frac{g_{n-2}(|z_1+B|)}
{|z_1+B|} \quad \quad \mu \neq 1. \nonumber
\end{eqnarray}
Therefore the matrix of second derivatives separates in $n$ $N\times N$ blocks.
For $\mu=1$ and for $\mu\ne 1$ the eigenvalues of the corresponding matrix are
given in terms of the eigenvalues $R_i$ of $R_{ij}$ by, respectively:
\begin{eqnarray}\label{eigen1}
-\frac{1}{2}R_i^{-1}+g'_{n-2}(|z_1+B|)&& \quad \quad \mu=1 \nonumber \\
-\frac{1}{2}R_i^{-1}+\frac{g_{n-2}(|z_1+B|)}{|z_1+B|}&& \quad \quad \mu \neq 1.
\end{eqnarray}
From the definition of $R_{ij}$ and from (\ref{spectrumrel}) we have the
following inequalities:
\begin{equation}\label{ineq1}
\frac{1}{\beta}\leq \frac{1}{2}R_i^{-1}\leq \frac{1}{\beta}\frac{\tilde{N}}
{\epsilon}.
\end{equation}
Therefore, the same analysis presented in Appendix C, with equations
(\ref{hesh.2}) and (\ref{hesh.3}), can be performed, with the conclusion that
the relevant stationary point is that with $\mathbf{z}$ parallel to
$\mathbf{h}$, while the other two solutions with $z_1<0$, present for
sufficiently small $h$ and $\beta>\beta_c=n$, are not maxima (except for $n=1$,
when one of the two is a local maximum).

For $h=0$, Eq. (\ref{solst3}), as already seen, determines only the modulus $z$
of the stationary point, becoming $z=\beta g_{n-2}(z)$, which has always a
solution $z=0$, and, for $\beta>\beta_c$, also a solution with positive $z$,
which is infinitely degenerate for $n>1$ and doubly degenerate for $n=1$.
In analogy with the $\alpha=0$ case, one should then perform in (\ref{Z1c}),
before applying the saddle point method, an angular integration over a global
rotation. This would give in (\ref{free1}) a further factor
$\frac{\ln \Omega_n}{N}$, which does not contribute in the thermodynamic
limit. In the point $z=0$ the hessian matrix is given by:
\begin{equation}\label{eigen2}
\left.\frac{\partial^2 f}{\partial z_{i\mu} \partial z_{j\nu}}\right|_{
\mathbf{z}=\mathbf{0}}=-\frac{1}{2}(R^{-1})_{ij}\delta_{\mu\nu} + \delta_{ij}
\delta_{\mu\nu}\frac{1}{n} \, ,
\end{equation}
with eigenvalues:
\begin{equation}\label{eigen3}
-\frac{1}{2}R_i^{-1}+\frac{1}{n} \, .
\end{equation}
Using (\ref{ineq1}), we see that for $\beta<\beta_c$ this stationary point is
a maximum, while for $\beta>\beta_c$ it is not a maximum. In the case of a
point $\mathbf{z}^*$ whose positive modulus $z^*$ satisfies the
self-consistency equation for $\beta>\beta_c$, we have:
\begin{eqnarray}\label{eigen4}
\left.\frac{\partial^2 f}{\partial z_{i\mu} \partial z_{j\nu}}\right|_{
\mathbf{z}=\mathbf{z}^*}&=& \delta_{\mu\nu}\left[-\frac{1}{2}(R^{-1})_{ij}
+\delta_{ij}\frac{1}{\beta}\right]
\nonumber \\ &+& \delta_{ij}\frac{z_\mu^*z_\nu^*}{z^{*2}}\left( g'_{n-2}(z^*)
-\frac{g_{n-2}(z^*)}{z^*}\right).
\end{eqnarray}
We can look at this matrix as being given by the sum of different
contributions, each one negative semi-definite. For the term multiplying
$\delta_{\mu\nu}$ this can be seen from (\ref{ineq1}), while for the term
multiplying $\delta_{ij}$ we can refer to the analysis performed in Appendix
C for expression (\ref{hesh0.2}). The degeneracy of the stationary point
is reflected in the existence of $n-1$ eigenvalues equal to zero; however,
we have just noted that in this case an integration over the region of
degeneracy is implied before the application of the saddle point. This
integration, that gives a vanishing contribution in the thermodynamic limit,
is performed exactly over the directions corresponding to the eigenvalues
equal to zero; as a consequence, these eigenvalues of
$\det \left(-\frac{1}{2}H_0\right)$ must not be taken into account in the
stability analysis.

\subsubsection{Possible non homogeneous solutions}
Now we consider other possible solutions of (\ref{solst2}), non homogeneous
on the lattice. We do not prove if they exist and if, in that case, they are
maxima; rather, we prove that, if they exist and are maxima, they are local,
i.e., the value of $f$ is in those points smaller than for our homogeneous
solution. Incidentally, we note that if a non homogeneous stationary point
exists, then, because of translational invariance, all the configurations,
obtained by the stationary one by any translation, are also stationary points
with the same value. Let us begin by rewriting the stationary point equations
(\ref{solst2}) in another form, posing $z_{i\mu}+B_\mu\equiv w_{i\mu}$:
\begin{eqnarray}\label{solst4}
w_{i\mu}&=&2 \sum_{j=1}^N R_{ij} g_{n-2}(|\mathbf{w}_j|)
\frac{w_{j\mu}}{|\mathbf{w}_j|}  + B_\mu \nonumber \\
&&\mu=1,\dots,n \quad \quad i=1,\ldots,N.
\end{eqnarray}
From this we can derive an equality, useful later, that is verified for our
homogeneous solution. Taking, as before and without loss of generality $B_\mu
=B\delta_{\mu 1}$, and the relevant homogeneous solution with $w_{i\mu}=
w\delta_{\mu 1}$ (i.e., $w_{i\mu}=0$ for $\mu=1$ and $w_{i1}=w\geq0$), we
obtain:
\begin{equation}\label{solst5}
w=\beta g_{n-2}(w)+B,
\end{equation}
where again we have used $2\sum_j R_{ij}=\beta$ for each $i$. This is valid
also for $B=0$. Let us now consider another possible solution of (\ref{solst4})
for which not all the $|\mathbf{w}_i|$s are equal. In this case we take the
first axis in the direction of the $\mathbf{w}_i$ with the largest modulus, be
it $\mathbf{w}_l$, and from (\ref{solst4}) we have:
\begin{eqnarray}\label{ineq2}
w_{l1}&=&|\mathbf{w}_l|=2\sum_{j=1}^N R_{lj}g_{n-2}(|\mathbf{w}_j|)
\frac{w_{j1}}{|\mathbf{w}_j|} + B_1 \nonumber \\
&\leq& 2\sum_{j=1}^N R_{lj}g_{n-2}(|\mathbf{w}_j|) + B \nonumber \\
&<& 2\sum_{j=1}^N R_{lj}g_{n-2}(|\mathbf{w}_l|) + B 
=\beta g_{n-2}(|\mathbf{w}_l|) + B,
\end{eqnarray}
where we have used the monotonicity of $g_{n-2}(x)$ and that $B_1 \le B$.
Therefore, a non homogeneous solution with different moduli for the
$\mathbf{w}_i$s is such that all these moduli are smaller than that of the
homogeneous solution satisfying (\ref{solst5}). In fact, the properties of the
functions $g$ (see also Fig. 1) imply that $|\mathbf{w}_l|$ in (\ref{ineq2})
is smaller than $w$ satisfying (\ref{solst5}). The same is true for a solution
with all equal moduli but different directions; in this case, in (\ref{ineq2})
the first inequality becomes strict and the second becomes an equality. It is
now sufficient, analogously to what we have done at the end of Appendix C for
the $\alpha=0$ case, to see that the exponent $f$ in (\ref{Z1c}) is an
increasing function of the moduli $|\mathbf{w}_i|$, and this will prove that
all non homogeneous solutions of (\ref{solst2}) are at most local maxima. It is
not difficult to show, using (\ref{solst4}), that, as a function of the
$\mathbf{w}_i$s, $f$ in the stationary points is given by:
\begin{equation}\label{fstat}
f=\sum_{i=1}^N\left[-\frac{1}{2}g_{n-2}(|\mathbf{w}_i|)|\mathbf{w}_i|+
\frac{1}{2}g_{n-2}(|\mathbf{w}_i|)\frac{\mathbf{w}_i\cdot\mathbf{B}}
{|\mathbf{w}_i|} + \ln G_{n-2}(|\mathbf{w}_i|)\right].
\end{equation}
Posing $\mathbf{w}_i=x_i\mathbf{s}_i$, with $x_i\equiv|\mathbf{w}_i|$ and
$\mathbf{s}_i$ unitary vectors, (\ref{fstat}) becomes:
\begin{equation}\label{fstat1}
f=\sum_{i=1}^N\left[-\frac{1}{2}g_{n-2}(x_i)x_i+
\frac{1}{2}g_{n-2}(x_i)\mathbf{s}_i\cdot\mathbf{B}
+ \ln G_{n-2}(x_i)\right].
\end{equation}
The differentiation with respect to $x_i$ gives, as it is easily seen:
\begin{equation}\label{fstder}
\frac{\partial f}{\partial x_i}=\frac{1}{2}x_i\left(\frac{g_{n-2}(x_i)}{x_i}
-g'_{n-2}(x_i)\right) +\frac{1}{2}g'_{n-2}(x_i)\mathbf{s}_i\cdot\mathbf{B}.
\end{equation}
The first term is always positive, according to property 5 of Appendix B, and
also $g'_{n-2}(x_i)$ is always positive. This proves that for the homogeneous
solution, satisfying (\ref{solst5}) and for which
$\mathbf{s}_i\cdot\mathbf{B}=B$ for each $i$, $f$ has the largest value.

\subsubsection{The free energy}
Now we are in the position to compute the terms in square brackets in
(\ref{free1}) and to write a more explicit expression for the specific free
energy $F$ in (\ref{free1}). We first write
down an expression for $\max f$. The maximum $\mathbf{z}^*$ is homogeneous,
then we have to know $\sum_{ij}(R^{-1})_{ij}$. Since $\sum_j R_{ij}=\frac
{\beta}{2}$ for each $i$, then $\sum_j (R^{-1})_{ij}=\frac{2}{\beta}$ for each
$i$. In fact, the first expression tells that an homogeneous vector is an
eigenvector of $R$ with eigenvalue $\frac{\beta}{2}$; therefore, the same
vector is an eigenvector of $R^{-1}$ with eigenvalue $\frac{2}{\beta}$, that
gives the second expression. It follows immediately that:
\begin{equation}\label{fmax}
\max_{\{z_{i\mu}\}} f=N\left[-\frac{1}{2\beta}\mathbf{z}^{*2} + \ln
G_{n-2}\left(|\mathbf{z}^* + \mathbf{B}|\right)\right].
\end{equation}

We now compute $\ln \det \left(-\frac{1}{2}H_0\right)$. Performing the
stability analysis we have explicitly considered the eigenvalues of the hessian
matrix in the homogeneous maximum. Summarizing the results, we can write the
expression below for the eigenvalues of $-\frac{1}{2}H_0$, valid both when the
maximum is at a positive value of $z^*$ (i.e., when $B>0$ and when $B=0$ but
$\beta>\beta_c=n$) and when it is at $z^*=0$. It is easily seen (always
considering, for convenience, $\mathbf{B}$ along the first axis, and
$\mathbf{z}^*$ along the first axis also when $B=0$ and $\beta>\beta_c$) that
the eigenvalues are given by:
\begin{eqnarray}\label{eigen5}
\frac{1}{4}R_i^{-1} - p_1(z^*)&& \nonumber \\
\frac{1}{4}R_i^{-1} - p_2(z^*)&& \quad \quad n-1 \quad {\rm times}
\end{eqnarray}
for $i=1,\dots,N$. Here, when $z^*>0$, $p_1(z^*)\equiv\frac{1}{2}g'_{n-2}
(z^*+B)$ and $p_2(z^*)\equiv\frac{1}{2\beta}\frac{z^*}{z^*+B}$; while, when
$z^*=0$, $p_1(z^*)=p_2(z^*)=\frac{1}{n}$. Therefore we have:
\begin{eqnarray}\label{deth01}
&&\det\left(-\frac{1}{2}H_0\right) = \prod_{i=1}^N\left[\left(\frac{1}{4}
R_i^{-1}-p_1(z^*)\right)\left(\frac{1}{4}R_i^{-1}-p_2(z^*)\right)^{n-1}\right]
= \nonumber \\
&=&\left(\frac{1}{4}\right)^{Nn}\left(\det R\right)^{-n}\prod_{i=1}^N\left[
\left(1-4R_ip_1(z^*)\right)\left(1-4R_ip_2(z^*)\right)^{n-1}\right].
\end{eqnarray}
The fact that we have to disregard the zero eigenvalues when $B=0$ and $\beta
>\beta_c$ is automatically taken into account in the above expression. We then
obtain:
\begin{eqnarray}\label{deth02}
&&-\frac{1}{2} \ln \det\left(-\frac{1}{2}H_0\right) = Nn \ln 2 +\frac{n}{2}
\ln \det R + \nonumber \\
&+&\sum_{i=1}^N\left[\ln\left(1-4R_ip_1(z^*)\right)+(n-1)\ln\left(
1-4R_ip_2(z^*)\right)\right].
\end{eqnarray}
We will show in a moment that the last sum does not contribute to
(\ref{free1}); then, using (\ref{fmax}) and (\ref{deth02}) in (\ref{free1}),
we have the following expression for the free energy:
\begin{equation}\label{free2}
-\beta F=-\frac{\beta}{2}+\ln \Omega_{n-1}-\frac{1}{2\beta}\mathbf{z}^{*2}
+\ln G_{n-2}\left(|\mathbf{z}^* +\mathbf{B}|\right)
\end{equation}
identical to (\ref{freemf}). We then obtain the same magnetization $\frac{
\mathbf{z}^*}{\beta}$ as for $\alpha=0$ (see also (\ref{solmagn})), and the
same equation of state (see (\ref{eqstmf}) and (\ref{soleqst})). Therefore,
we still have a second order phase transition at $\beta=\beta_c$, where $M$
becomes $0$. Of course we
can make the same comments as for the mean field case (see the last paragraph 
of Section \ref{MeanField}), concerning the addition of the kinetic part.

This concludes, apart from the remaining technical points treated below, our
proof of the universality of all the models, for each $n$, when $\alpha < d$.
We repeat that the difference, when $n$ varies, is in the value of the
critical temperature $T_c=\frac{1}{n}$, but the overall behaviour is the same
for all values of $n$.

We are left with two points: the proof that the sum in (\ref{deth02}) does
not contribute to (\ref{free1}), and the justification of the saddle point
method. The next subsection is dedicated to the second point; here we treat the
first. From (\ref{spectrumrel}) we have:
\begin{equation}\label{specr1}
\frac{\beta}{2}\frac{\epsilon}{\tilde{N}}\leq R_i \leq \frac{\beta}{2} \, ,
\end{equation}
with the important specification that only a vanishing fraction of these
eigenvalues remains finite in the thermodynamic limit. We can therefore write:
\begin{eqnarray}\label{deth03}
&&\lim_{N\rightarrow \infty}\frac{1}{N}\sum_{i=1}^N\left[\ln\left(
1-4R_ip_1(z^*)\right)+(n-1)\ln\left(1-4R_ip_2(z^*)\right)\right]=\nonumber \\
&=&\lim_{N\rightarrow \infty}\left\{-\frac{1}{N}\sum\: '\left[4R_i\left(
p_1(z^*)+(n-1)p_2(z^*)\right)\right]\right. \\
&&\left. +\frac{1}{N}\sum\, '' \left[\ln\left(
1-4R_ip_1(z^*)\right)+(n-1)\ln\left(1-4R_ip_2(z^*)\right)\right]\right\},
\nonumber
\end{eqnarray}
where the first sum is on the vanishing eigenvalues, and the second on the
others; in the first we have substituted the logarithm with its first order
approximation. We also point out that, following our stability analysis, the
arguments of the logarithms in (\ref{deth03}) are between $0$ and $1$. We can
indicate with $\eta$ the fraction of eigenvalues present
in the second sum; if we also denote with $R'_{max}$ and $R''_{max}$ the
largest eigenvalues in the two sums, the above positive expression is bounded
from above by:
\begin{equation}\label{deth04}
(1-\eta)R'_{max}+\eta|\ln (1-R''_{max})|.
\end{equation}
In the thermodynamic limit this expression goes to zero, since both $R'_{max}$
and $\eta$ go to zero. Therefore, the sum in (\ref{deth02}) does not
contribute to the specific free energy (\ref{free1}).

\subsection{Justification of the saddle point}
In the integral (\ref{Z1c}) the number of integration variables itself grows
with $N$. The value of the maximum of the exponent diverges with $N$ (this is
seen in (\ref{fmax})); but the use of the saddle point gives a proper
evaluation of the integral if also the curvatures in all directions diverge
with $N$, i.e., if all the eigenvalues of $-\frac{1}{2}H_0$ diverge with $N$.
We now show that this is not exactly verified, but since what we are
interested in is the evaluation of the specific free energy $F$ (see 
(\ref{free1}) and (\ref{free2})), we also show that nevertheless the saddle
point method is still justified.

The eigenvalues of $-\frac{1}{2}H_0$ are given by (\ref{eigen5}). The
functions $p_1(z^*)$ and $p_2(z^*)$ are finite, and for $\frac{1}{4}R_i^{-1}$
we can look at expression (\ref{ineq1}). According to our previous analysis in
subsection \ref{intermat}, among the values of $\frac{1}{4}R_i^{-1}$ only a
fraction, vanishingly small in the thermodynamic limit, does not diverge with
$N$. Along the directions corresponding to those eigenvalues the integral
should be computed explicitly, reserving the saddle point method to the other
directions; however, we show that the error introduced using altogether the
saddle point vanishes in the computation of $F$.

Let us call collectively $\delta_l$ the eigenvalues of $-\frac{1}{2}H_0$, where
$l$ runs from $1$ to $P\equiv nN$, and indicate with $v_1,\dots,v_P$ the
integration variables in (\ref{Z1c}); then, the fact that the integral in that
expression is evaluated with the saddle point is equivalent the the following
replacement:
\begin{equation}\label{repl1}
\int d^Pv e^f \longrightarrow e^{\max f}\int d^Pv e^{-\sum_{l=1}^P
\delta_l v_l^2} \, .
\end{equation}
For the $v_l$s for which $\delta_l$ does not diverge with $N$ this replacement
is not a good approximation, and we should more correctly write something
like:
\begin{equation}\label{repl2}
e^{\max f}\int d^Pv e^{-\sum_{l=1}^{\bar{P}}\delta_l v_l^2 - u(v_{\bar{P}+1},
\dots,v_P)}
\end{equation}
with $u>0$ always, except that $u=0$ when $v_{\bar{P}+1}=\dots=v_P=0$, and
where $\delta_{\bar{P}+1},\dots,\delta_P$ (with $\bar{P}<P$) are the
$\delta_l$s that do not diverge with $N$. The previous expression can also be
written as:
\begin{equation}\label{repl3}
e^{\max f}\int d^Pv e^{-\sum_{l=1}^P\delta_l v_l^2}\frac{\int dv_{\bar{P}+1}
\dots dv_P e^{- u(v_{\bar{P}+1},\dots,v_P)}}{\int dv_{\bar{P}+1}\dots dv_P e^
{-\sum_{l=\bar{P}+1}^P\delta_l v_l^2}} \, .
\end{equation}
We are interested in the limit, when $N\rightarrow \infty$, of $\frac{1}{N}$
times the logarithm of this expression. We have already seen that, when
$N\rightarrow \infty$, $\frac{P-\bar{P}}{P}$ goes to zero. This implies that
the contribution of the last fraction in (\ref{repl3}) to the evaluation of the
free energy $F$ vanishes in the thermodynamic limit, and this is equivalent to
using the saddle point expression (\ref{repl1}).

\subsection{Microcanonical solution}\label{micro}
We briefly treat the point of the microcanonical solution of our system.
The equivalence of different ensembles is a problem of general character, and
here we only show that for our system the canonical and microcanonical
ensembles are equivalent.

For a generic system, indicating collectively with $\Gamma$ the coordinates of
its phase space, its canonical partition function can be written as:
\begin{equation}\label{canon}
Z(\beta)=\int d\Gamma \: e^{-\beta H(\Gamma)}=\int_0^\infty dE \: \omega(E)
e^{-\beta E}=\int_0^\infty dE \: e^{\left[-\beta E + \ln \omega(E)\right]} \, ,
\end{equation}
where $\omega (E)$ is the microcanonical density of states:
\begin{equation}\label{microc}
\omega(E)=\int d\Gamma \: \delta\left(H(\Gamma)-E\right).
\end{equation}
In (\ref{canon}) we have supposed that the hamiltonian is bounded from below
at $0$ (as in our case) without loss of generality; the dependence on the
number $N$ of particles is not explicitly written. Introducing the specific
energy $U=\frac{E}{N}$ and using the definition of the microcanonical specific
entropy in the thermodynamic limit
\begin{equation}\label{microent}
S(U)=\lim_{N\rightarrow\infty}\frac{1}{N}\ln \omega(E)
\end{equation}
then, from the last expression in (\ref{canon}) we see that the canonical
partition function can be computed, in the thermodynamic limit, by the saddle
point method, and the specific free energy is therefore given by:
\begin{equation}\label{canonfree1}
-\beta F(\beta)=\lim_{N\rightarrow\infty}\frac{1}{N}\ln Z(\beta)
=\max_U\left[ -\beta U + S(U)\right].
\end{equation}
If $S(U)$ is convex, i.e., if $\frac{d^2 S}{dU^2}<0$, this relation defines
a single value of $U$ for each $\beta$, $U_{mc}(\beta)$, given by
$\left.\frac{dS}{dU}\right|_{U=U_{mc}}=\beta$, and we can write:
\begin{equation}\label{canonfree2}
-\beta F(\beta)=-\beta U_{mc}(\beta) +S(U_{mc}(\beta)).
\end{equation}
On the other hand, if $S(U)$ has a concavity region, as in the presence of
first order phase transitions, then it can be easily deduced from
(\ref{canonfree1}) that the
temperature derivative of $F$ has a discontinuity. In our system we have only
a second order transition, and in fact we have no discontinuity in
$\frac{\partial F}{\partial \beta}$; therefore $S(U)$ is convex and
(\ref{canonfree2}) holds
(of course this is true whether or not we consider the trivial kinetic energy
contribution to $F$). It is now easy to show the equivalence of the two
ensembles. In fact the entropy $S_c(\beta)$ computed from the canonical
ensemble is obtained from (\ref{canonfree2}) as:
\begin{eqnarray}\label{canonent}
S_c(\beta)&=&-\frac{\partial F}{\partial T}=\beta^2\frac{\partial F}
{\partial \beta} =
\beta^2\frac{\partial}{\partial \beta}\left[U_{mc}(\beta)-\frac{1}{\beta}
S(U_{mc}(\beta))\right]= \nonumber \\
&=&\beta^2\left[\frac{dU_{mc}}{d\beta}+\frac{1}{\beta^2}S(U_{mc}(\beta))-
\frac{1}{\beta}\left.\frac{dS}{dU}\right|_{U=U_{mc}}\cdot\frac{dU_{mc}}{d\beta}
\right]= \nonumber \\
&&=S(U_{mc}(\beta))
\end{eqnarray}
using that $\left.\frac{dS}{dU}\right|_{U=U_{mc}}=\beta$. Also for the
canonical energy $U(\beta)$ we obtain:
\begin{eqnarray}\label{canonene}
U(\beta)&=&\frac{\partial}{\partial \beta}\left(\beta F(\beta)\right)=
\frac{\partial}{\partial \beta}\left[\beta U_{mc}(\beta)-
S(U_{mc}(\beta))\right]= \nonumber \\
&=&U_{mc}(\beta)+\beta\frac{dU_{mc}}{d\beta}-
\left.\frac{dS}{dU}\right|_{U=U_{mc}}\cdot\frac{dU_{mc}}{d\beta}=U_{mc}(\beta).
\end{eqnarray}

\section{Discussion and conclusions}\label{discuss}
In this paper we have considered the problem of computing the partition
function of lattice magnetic models with long-range couplings. We have studied
a class of models in which the decay of the interaction with distance is
gauged by the exponent $\alpha$, smaller than the spatial dimension $d$ in
which the lattice is embedded. From the technical point of view, our study
has required several steps: {\it i)} the analysis of the spectrum of the matrix
$R$ for a consistent application of the well known gaussian identity sometimes
called Hubbard-Stratonovich transform (see (\ref{gauss}) and (\ref{gausstrb}));
{\it ii)} the analysis of a class of functions
related to the modified Bessel functions of the first kind, and characterized
by the index $n$, the dimension of the spins; {\it iii)} the application of the
saddle point method to an integral with a diverging number of integration
variables, which had to be justified; {\it iv)} the proof that possible
stationary points, if any, in the general case (in the strict sense $0<\alpha
<d$), of the exponent in (\ref{Z1c}) can be at most only local maxima, and are
therefore irrelevant in the thermodynamic limit.

In our computations, we have
not explicitly considered the case $\alpha=d$ and we have restricted the
long-range couplings to a power form. However, it is not difficult to argue
that also for a power decay with $\alpha=d$ and for a more generic form of
$J_{ij}$ with a long-range character (i.e., $\sum_{j=1}^N |J_{ij}|$ diverging
with $N$) we would have obtained the same results. In fact, the
basic points for our computations are: firstly, the divergence with $N$ of the
quantity $S$ in (\ref{lambdaest1}) and consequently of $\tilde{N}$ in
(\ref{ntilde1}), and secondly the fact that only a vanishing fraction, in the
thermodynamic limit, of the eigenvalues $\lambda_{\mathbf{k}}$ diverges, with
the consequence, given the first point, that only a vaninshing fraction of
$\frac{\lambda_{\mathbf{k}}}{\tilde{N}}$ (see (\ref{spectrumrel})) remains
finite. When $\alpha=d$ all these points would remain; the behaviour of the
quantity $S$ in (\ref{lambdaest1}) for large $N$ would then be given by
$\ln N$. Also for a generic $J_{ij}$ all these points would still be true, even
if the divergence law of $S$ with $N$ could possibly be difficult to write
explicitly.

Let us remark that we have also shown the equivalence, for these models, of
the microcanonical and the canonical ensembles, in spite of the lack of
additivity. 

In conclusion, we have found an entire class of lattice spin models with a
universal thermodynamic behaviour.

These results on lattice systems can be helpful for the thermodynamics of
continuous long-range systems; it should be stressed, however, that, without
the Kac's prescription, not only additivity, but also extensivity is violated,
and this could imply ensemble inequivalence. The microcanonical ensemble
would then be the natural framework for the study of those cases.

\section*{Appendix A: The functions $G_p(x)$ and $g_p(x)$}\label{Apdxa}
In (\ref{MF.Z2}) we have, for each site $i$, an integral of the form:
\begin{equation}\label{unitsph1}
\int d\Theta \: e^{\mathbf{S} \cdot \mathbf{z}}
\end{equation}
extended on the surface of the unit sphere in dimension $n\geq 2$, where the
unit vector $\mathbf{S}$ lies. For $n=1$ the integral is substituted by the
sum over $S=\pm 1$; in that case also $\mathbf{z}$ has a single component
$z$, (\ref{unitsph1}) is given by $2\cosh(z)$, and we do not need further
analysis. For $n\geq 2$ we can choose the axes of the unit sphere such
that $\mathbf{z}$ lies along one of them; besides, we can introduce polar
coordinates in $n$ dimensions, taking as the polar axis the one along
$\mathbf{z}$. Then, if $z=|\mathbf{z}|$, it is easy to show that
(\ref{unitsph1}) becomes:
\begin{equation}\label{unitsph2}
\int d\Theta \: e^{\mathbf{S} \cdot \mathbf{z}}
=\Omega_{n-1} \int_0^\pi d\theta \sin^{n-2}\theta e^{z\cos\theta},
\end{equation}
where $\theta$ is the polar angle, while $\Omega_n$ is the area of the unit
sphere in $n$ dimensions (with $\Omega_1=2$); it can be expressed in terms of
the gamma function as $\Omega_n=2\pi^{n/2}/\Gamma(n/2)$. This last factor
is not considered further, and for the integral in (\ref{unitsph2}) we
introduce the following notation:
\begin{equation}\label{GdefA}
G_p(x)=\int_0^\pi d\theta \sin^p\theta e^{x\cos\theta}=
2\int_0^{\pi/2} d\theta \sin^p\theta \cosh(x\cos\theta),
\end{equation}
where the parameter $p$ is related to the dimension $n\ge 2$ of the spin
vector by $p=n-2$, and therefore $p$ takes the non negative integer values.
If $n=1$ it is understood in Section \ref{MeanField}
and Section \ref{Solution} that $G_{-1}(x)=\cosh(x)$ and $\Omega_0=2$. It
can be shown (see next appendix) that $G_p(x)$ is related to
the modified Bessel function of first kind of order $\frac{p}{2}$, i.e.
\begin{equation}\label{relat}
G_p(x) \propto \frac{I_{\frac{p}{2}}(x)}{x^{\frac{p}{2}}},
\end{equation}
relation that is valid also for $p=-1$.
The other functions of interest are the derivative of $G_p(x)$:
\begin{equation}\label{GpdefA}
G'_p(x)=\int_0^\pi d\theta \cos\theta \sin^p\theta e^{x\cos\theta}=
2\int_0^{\pi/2} d\theta \cos\theta \sin^p\theta \sinh(x\cos\theta)
\end{equation}
and the logarithmic derivative:
\begin{eqnarray}\label{gdefA}
g_p(x)=\frac{d}{dx}\ln G_p(x)&=&\frac{G'_p(x)}{G_p(x)}=
\frac{\int_0^\pi d\theta \cos\theta \sin^p\theta e^{x\cos\theta}}
     {\int_0^\pi d\theta \sin^p\theta e^{x\cos\theta}}= \nonumber \\
&=&\frac{\int_0^{\pi/2} d\theta \cos\theta \sin^p\theta \sinh(x\cos\theta)}
     {\int_0^{\pi/2} d\theta \sin^p\theta \cosh(x\cos\theta)}.
\end{eqnarray}
For $n=1$ (i.e., $p=-1$) they are substituted by $\sinh(x)$ and by $\tanh(x)$,
respectively. In the next appendix we prove some properties of the functions
$G_p(x)$, $G'_p(x)$ and $g_p(x)$, which are necessary for the analysis of the
self-consistency equations (\ref{SC}) or (\ref{solst3}).

\section*{Appendix B: The properties of $g_p(x)$}\label{Apdxb}
As we have seen, in (\ref{GdefA}), (\ref{GpdefA}) and (\ref{gdefA}) the
parameter $p$ takes the non negative integer values. We note however that, if
we take $p$ to be any real number, the functions inside the integrals are
integrable for any $p$ greater than $-1$, and besides it can be easily proved
that the limit for $p\rightarrow -1$ of (\ref{gdefA}) is exactly $\tanh(x)$.
As a consequence all the properties that we will show
for $G_p(x)$, $G'_p(x)$ and $g_p(x)$ are valid for any $p \geq -1$ (if
necessary, for $p=-1$ the explicit expressions that we have given can be
invoked). Some properties will be
immediately evident from the definitions, while others,
especially those of $g_p(x)$, will need a proof.

Considering first $G_p(x)$ and $G'_p(x)$, we begin with their properties of
symmetry with respect
to $x$ inversion and monotonicity. In particular, the positive function
$G_p(x)$ is even in $x$ and is monotonically increasing for $x>0$. This is most
easily seen from the second expression in (\ref{GdefA}). From the second
expression in (\ref{GpdefA}) it is also readily seen that the odd function
$G'_p(x)$, positive for $x>0$ and negative for $x<0$, is monotonically
increasing. A relation which is useful for the following can be derived with
an integration by parts in (\ref{GpdefA}), namely:
\begin{equation}\label{GprG}
G'_p(x)=\frac{x}{p+1}G_{p+2}(x).
\end{equation}
We also give the value of the first few derivatives of $G_p(x)$ in $x=0$.
Since $G_p(x)$ is even, the odd derivatives vanish for $x=0$, while
from the expression
\begin{equation}\label{Gderm}
G^{(m)}_p(x)=\frac{d^m}{dx^m}G_p(x)=\int_0^\pi d\theta \cos^m\theta
\sin^p\theta e^{x\cos\theta}
\end{equation}
integrations by parts, after posing $x=0$, give, for the second and fourth 
derivatives in $x=0$, the expressions:
\begin{equation}\label{Gsecfou}
G''_p(0)=\frac{1}{p+2}G_p(0), \quad \quad G^{(4)}_p(0)=\frac{3}
{(p+4)(p+2)}G_p(0).
\end{equation}
These are useful for the computations of the necessary derivatives
of $g_p(x)$ in $x=0$).
Before turning to $g_p(x)$, it is interesting to derive a second order
differential equation satisfied by $G_p(x)$. From the relation
$G''_p(x)=G_p(x)-G_{p+2}(x)$, obtainable from (\ref{Gderm}) posing
$\cos^2\theta=1-\sin^2\theta$, using (\ref{GprG}) we have:
\begin{equation}\label{besgen}
x^2 G''_p(x)+(p+1)x G'_p(x)-x^2 G_p(x)=0.
\end{equation}
From this it is easy to derive the equation satisfied by $W_p(x)\equiv
{x^{\frac{p}{2}}}G_p(x)$:
\begin{equation}\label{eqbes}
x^2 W''_p(x)+ x W'_p(x)-\left[x^2+\left(\frac{p}{2}\right)^2\right] W_p(x)=0,
\end{equation}
which is the modified Bessel equation with parameter $\frac{p}{2}$. Knowing
also the limiting properties, for $x \rightarrow 0$, of $G_p(x)$, it can be
concluded that $W_p(x)$ is proportional to the modified Bessel function
of first kind of order $\frac{p}{2}$, $I_{\frac{p}{2}}(x)$.

Regarding $g_p(x)$, we show the following properties:
\begin{enumerate}
\item{$g_p(x)$ is odd, positive for $x>0$ and negative for $x<0$;}
\item{$g_p(x)$ is, in modulus, smaller than $1$, and it tends to $\pm 1$
for $x \rightarrow \pm \infty$;}
\item{$g'_p(x)$, which is even, is positive;}
\item{$\frac{g_p(x)}{x}$, which is even and positive, has a derivative which
is negative for $x>0$ and positive for $x<0$;}
\item{$\frac{g_p(x)}{x}-g'_p(x)$, which is even, is positive;}
\item{$g''_p(x)$, which is odd, is negative for $x>0$ and positive for $x<0$.}
\end{enumerate}
We give the proofs of these properties.

{\it Property 1}.
It is an immediate consequence of the definition and of the simmetry
and monotonicity properties of $G_p(x)$ and $G'_p(x)$.

Although also the second can be easily derived from the definition, we prove
it, together with the remaining, by introducing the first order differential
equation satisfied by $g_p(x)$. From the definition of $g_p(x)$ we have:
\begin{equation}\label{eqgp1}
g'_p(x)=\frac{G''_p(x)}{G_p(x)}-\left(\frac{G'_p(x)}{G_p(x)}\right)^2=
\frac{G''_p(x)}{G_p(x)}-g^2_p(x).
\end{equation}
Substituting $G''_p(x)$ as a function of $G'_p(x)$ and $G_p(x)$ from
(\ref{besgen}), we have:
\begin{equation}\label{eqgpd}
g'_p(x)=1-\frac{p+1}{x}g_p(x)-g^2_p(x).
\end{equation}
This equation must be supplied with the initial condition $g_p(0)=0$. The
first few derivatives of $g_p(x)$ in $x=0$ can be obtained from (\ref{eqgpd})
or from the derivatives of $G_p(x)$. In particular, since $g_p(x)$ is odd,
the even derivatives in $x=0$ vanish, while for the first and third derivatives
we have:
\begin{equation}\label{gfirthi}
g'_p(0)=\frac{1}{p+2}, \quad \quad g'''_p(0)=-\frac{6}{(p+4)(p+2)^2}.
\end{equation}
We now proceed with the remaining proofs, using (\ref{eqgpd}) and
(\ref{gfirthi}).

{\it Property 2, first part}. $g_p(x)$ starts at $x=0$ with positive derivative
$\frac{1}{p+2}$ (see (\ref{gfirthi})). Therefore, if, for $x>0$, it intersects
the value $g_p=1$, it must do that at least once with non negative derivative.
But from (\ref{eqgpd}) we get for $g_p=1$ that $g'_p=-\frac{p+1}{x}$, which is
negative for finite $x>0$. Then, $g_p(x)$ remains, for $x>0$, between $0$ and
$1$.

{\it Property 3}. By differentiating (\ref{eqgpd}), we get the following
equation for $g''_p(x)$:
\begin{equation}\label{eqgpd2}
g''_p(x)=\frac{p+1}{x^2}g_p(x)-\frac{p+1}{x}g'_p(x)-2g_p(x)g'_p(x).
\end{equation}
Since $g'_p(0)=\frac{1}{p+2}>0$, if $g'_p(x)$ intersects the value $0$, it
must do that at least once with $g''_p$ non positive for $x>0$ and non negative
for $x<0$. But from (\ref{eqgpd2}) we get for $g'_p=0$ that $g''_p=
\frac{p+1}{x^2}g_p$, which is positive for finite $x>0$ and negative for
finite $x<0$. Then, $g'_p(x)>0$ for all $x$.

{\it Property 2, second part}. We restrict to positive $x$; the proof for
negative $x$ follows from the fact that $g_p(x)$ is odd. Since $g'_p(x)>0$ and
$g_p(x)<1$ for finite $x$, $g_p(x)$ must have an asymptote, and for
$x \rightarrow \infty$ $g'_p(x)$ must tend to $0$. In that limit (\ref{eqgpd})
becomes $0=1-g_p^2$. Therefore, for $x \rightarrow \infty$ $g_p(x)$ tends to
$1$.

{\it Property 4}. We define $h_p(x)=\frac{g_p(x)}{x}$. From
(\ref{eqgpd}) it is easy to derive the differential equation satisfied
by $h_p(x)$:
\begin{equation}\label{eqhp}
h'_p(x)=\frac{1}{x}-\frac{p+2}{x}h_p(x)-xh^2_p(x).
\end{equation}
Following what has already been proved, $h_p(x)$ is even and positive,
and we have:
\begin{equation}\label{hder}
h_p(0)=\frac{1}{p+2}, \quad \quad h'_p(0)=0, \quad \quad h''_p(0)=
-\frac{2}{(p+4)(p+2)^2}.
\end{equation}
By differentiating (\ref{eqhp}) we have the equation:
\begin{equation}\label{eqhp2}
h''_p(x)=-\frac{1}{x^2}+\frac{p+2}{x^2}h_p(x)-\frac{p+2}{x}h'_p(x)
-h^2_p(x)-2xh_p(x)h'_p(x).
\end{equation}
From (\ref{hder}) we see that, for $x$ positive sufficiently small,
$h'_p(x)$ is negative; therefore, if for larger $x$ $h'_p(x)$ becomes equal
to $0$, it must do that with a non negative $h''_p(x)$, and with $h_p(x)$
smaller than $\frac{1}{p+2}$. But from (\ref{eqhp2}) we have for $h'_p(x)=0$
that $h''_p(x)=-h^2_p(x)-\frac{1}{x^2}\left[1-(p+2)h_p(x)\right]$, which is
negative. Then, for $x>0$ $h'_p(x)$ is always negative, and it can also be
easily seen that it tends to zero, together with $h_p(x)$, for $x \rightarrow
\infty$. The proof for negative $x$ follows from the fact that $h_p(x)$ is
even.

{\it Property 5}. This is a simple consequence of the previous property, once
we note that $h'_p(x)=\frac{g'_p(x)}{x}-\frac{g_p(x)}{x^2}=
\frac{1}{x}\left(g'_p(x)-\frac{g_p(x)}{x}\right)<0$.

{\it Property 6}. By differentiating (\ref{eqgpd2}), we get the following
equation for $g'''_p(x)$:
\begin{eqnarray}\label{eqgpd3}
g'''_p(x)&=&-2\frac{p+1}{x^3}g_p(x)+2\frac{p+1}{x^2}g'_p(x) + \nonumber \\
&-&\frac{p+1}{x}g''_p(x)-2g'^2_p(x)-2g_p(x)g''_p(x).
\end{eqnarray}
The positive function $g'_p(x)$ starts at $x=0$ with the value $\frac{1}{p+2}$,
and for sufficiently small $x$ $g''_p(x)$ is negative (see (\ref{gfirthi}));
therefore, if for larger $x$ $g''_p(x)$ becomes equal to $0$, it must do that
with a non negative $g'''_p(x)$. But from (\ref{eqgpd3}) we have for $g''_p(x)
=0$ that $g'''_p(x)=-2g'^2_p(x)-2\frac{p+1}{x^2}\left(\frac{g_p(x)}{x}
-g'_p(x)\right)$, which is negative, using property 5. Thus, for $x>0$,
$g''_p(x)$ is negative. The proof for negative $x$ follows from the fact
that $g''_p(x)$ is odd.

\section*{Appendix C: Mean field solution of the $\alpha =0$ case}\label{Apdxc}
We give here some details of the mean field solution. We start from the
stationary point equations (\ref{mfsol1t}), that we rewrite here:
\begin{equation}\label{mfsol1}
\frac{\partial f}{\partial z_\mu}=-\frac{1}{\beta}z_\mu + g_{n-2}
(|\mathbf{z}+\mathbf{B}|)\frac{z_\mu+B_\mu}{|\mathbf{z}+\mathbf{B}|}=0,
\quad \quad \mu=1,2,\dots,n.
\end{equation}
Let us begin with the case $h>0$ (we remind that $B=\beta h$). Without
loss of generality, we can take the magnetic field $\mathbf{h}$ in the
positive direction of the first axis, i.e., $h_1=h$ and $h_\mu=0$ for
$\mu \neq 1$. Then, from (\ref{mfsol1}) we have, for $\mu \neq 1$:
\begin{equation}\label{mfsol2}
z_\mu =\beta g_{n-2}
(|\mathbf{z}+\mathbf{B}|)\frac{z_\mu}{|\mathbf{z}+\mathbf{B}|},
\quad \quad \mu=2,\dots,n.
\end{equation}
The possible solutions of these $n-1$ equations are: $z_\mu=0$ for
each $\mu=2,\dots,n$, or, if some $z_\mu$ is not zero, such that
$\beta \frac{g_{n-2}
(|\mathbf{z}+\mathbf{B}|)}{|\mathbf{z}+\mathbf{B}|}=1$. But in the
second case the equation for $\mu=1$ would become $z_1=z_1+B$, which is not
acceptable since $B>0$. Thus, $z_\mu=0$ for $\mu >1$, and the remaining
equation for $\mu=1$ becomes:
\begin{equation}\label{mfsol3}
z_1=\beta g_{n-2}(|z_1+B|)\frac{z_1+B}{|z_1+B|}.
\end{equation}
From the symmetry properties of $g_n(x)$, described in the previous Appendix,
this equation is equivalent to:
\begin{equation}\label{mfsol4}
z_1=\beta g_{n-2}(z_1+B).
\end{equation}
This equation can be solved graphically. We use the properties of the
functions $g_n(x)$ that we have proved, and for a visual aid we refer to
Fig. 1. In particular, since the maximum of
the positive function $g'_{n-2}(x)$ is for $x=0$ and is equal to $\frac{1}{n}$,
and the odd function $g''_{n-2}(x)$ is negative for $x>0$, we have that
(\ref{mfsol4}) always admits a single positive solution for $z_1$, while,
if $\beta>\beta_c=n$, it can also have, for sufficiently small $h$, two
other negative solutions. From the stability analysis we will see that the
relevant solution is the positive one.

We now turn to the case $h=0$. The stationary point equations become:
\begin{equation}\label{mfsol5}
\frac{\partial f}{\partial z_\mu}=-\frac{1}{\beta}z_\mu + g_{n-2}
(z)\frac{z_\mu}{z}=0,
\quad \quad \mu=1,2,\dots,n
\end{equation}
with $z=|\mathbf{z}|$. It is readily seen that these equation determine
only the modulus $z$. In fact, posing $z_\mu=zs_\mu$, we see that the
unit vector $\mathbf{s}$, giving the direction of $\mathbf{z}$, is left
free, and the equation for $z$ is:
\begin{equation}\label{mfsol6}
z =\beta g_{n-2}(z),
\end{equation}
which is of the same form of (\ref{mfsol4}). Again, from the properties of
$g_n(x)$ and looking at Fig. 1, we have that for $\beta \leq \beta_c=n$ the
only solution is
$z=0$, while for $\beta>\beta_c$ we also have a positive solution. The
stability analysis will show that in this last case the relevant solution is
the positive one. To summarize, in both cases, with or without magnetic field,
the relevant stationary point $\mathbf{z}^*$ is such that its modulus $z^*$
satisfies the self-consistency equation (\ref{SC}), with the further
characteristics specified soon after it.

Let us now show the stability analysis. The second order derivatives of
$f(\mathbf{z})$ are given by:
\begin{eqnarray}\label{hesgen}
&&\frac{\partial^2 f}{\partial z_\mu \partial z_\nu}=-\frac{1}{\beta}
\delta_{\mu\nu}+g'_{n-2}(|\mathbf{z}+\mathbf{B}|)\frac{(z_\mu+B_\mu)
(z_\nu+B_\nu)}{|\mathbf{z}+\mathbf{B}|^2}\\ &-&g_{n-2}(|\mathbf{z}+\mathbf{B}|)
\frac{(z_\mu+B_\mu)(z_\nu+B_\nu)}{|\mathbf{z}+\mathbf{B}|^3}+\delta_{\mu\nu}
g_{n-2}(|\mathbf{z}+\mathbf{B}|)\frac{1}{|\mathbf{z}+\mathbf{B}|}. \nonumber
\end{eqnarray}
They form the hessian matrix, that, at the stationary point, is
negative definite if this point is a maximum.

Let us begin with the case
without magnetic field, $h=0$. In this case the second derivatives become:
\begin{eqnarray}\label{hesgenh0}
\frac{\partial^2 f}{\partial z_\mu \partial z_\nu}&=&-\frac{1}{\beta}
\delta_{\mu\nu} +g'_{n-2}(z)\frac{z_\mu z_\nu}{z^2} \nonumber \\ &-&g_{n-2}(z)
\frac{z_\mu z_\nu}{z^3}+\delta_{\mu\nu}g_{n-2}(z)\frac{1}{z}.
\end{eqnarray}
We have shown before that there is always a solution $\mathbf{z}=\mathbf{0}$.
Using the properties of the function $g_n(x)$, the hessian matrix at this
point is given by
\begin{equation}\label{hesh0.1}
\left.\left(\frac{\partial^2 f}{\partial z_\mu \partial z_\nu}\right)
\right|_{\mathbf{z}=\mathbf{0}}=\delta_{\mu\nu}\left(
\frac{1}{n}-\frac{1}{\beta}\right)
\end{equation}
and is already in diagonal form. We immediately see that above the critical
temperature (i.e., for $\beta<\beta_c=n$) the only stationary point
$\mathbf{z}=\mathbf{0}$ is a maximum; for $\beta>\beta_c$ it is a minimum and
therefore it is not the right solution. At the critical temperature $\beta=n$
the second derivatives vanish and the saddle point in (\ref{MF.Z3}) must be
solved considering derivatives of higher order, showing again that
$\mathbf{z}=\mathbf{0}$ is a maximum. We do not show it here. For the solution
with positive $z$ for $\beta>\beta_c$ (that has an undetermined direction if
$n>1$ and that we call $\mathbf{z}^*$), using the stationary point equation
(\ref{mfsol6}) we obtain for the hessian matrix:
\begin{equation}\label{hesh0.2}
\left.\left(\frac{\partial^2 f}{\partial z_\mu \partial z_\nu}\right)
\right|_{\mathbf{z}=\mathbf{z}^*}=\frac{z_\mu^* z_\nu^*}{z^{*2}}\left(
g'_{n-2}(z^*)-\frac{g_{n-2}(z^*)}{z^*}\right).
\end{equation}
The term in brackets is negative, according to the properties of the function
$g_n(x)$, and we only have to study the matrix $A_{\mu\nu}=z_\mu^* z_\nu^*$.
It is very easy to show that this matrix has an eigenvalue equal to $z^{*2}$
and $n-1$ eigenvalues equal to $0$. Therefore, coherently with the fact that
for $h=0$ the stationary point is infinitely degenerate if $n>1$, we find
that this point is a maximum, but if $n>1$ there are $n-1$ directions along
which $f(\mathbf{z})$ does not change.

We now consider the case with a magnetic field. We have studied the stationary
points with $h_1=h$ and $h_\mu=0$ for $\mu \neq 1$, for which $z_\mu =0$ for
$\mu \neq 1$. In general, for any $\mathbf{z}$ such that $z_\mu=0$ for
$\mu \ne 1$ (and $B_\mu=B\delta_{\mu 1}$), the hessian matrix becomes:
\begin{eqnarray}\label{hesh.1}
&&\left.\frac{\partial^2 f}{\partial z_\mu \partial z_\nu}\right|_{z_\sigma
=z_1\delta_{\sigma 1}}=\delta_{\mu\nu}\Bigl(-\frac{1}
{\beta}+g'_{n-2}(|z_1+B|)\frac{(z_\mu+B)^2}
{|z_1+B|^2}\delta_{\mu 1} \nonumber \\ &-&g_{n-2}(|z_1+B|)\frac{(z_\mu+B)^2}
{|z_1+B|^3}\delta_{\mu 1} + g_{n-2}(|z_1+B|)\frac{1}{|z_1+B|}\Bigr)
\end{eqnarray}
which is again diagonal. If we consider first the diagonal term with $\mu=1$,
we see that it is equal to
\begin{equation}\label{hesh.2}
\left.\frac{\partial^2 f}{\partial z_1^2}\right|_{z_\mu=z_1\delta_{\mu 1}}=
-\frac{1}{\beta}+g'_{n-2}(|z_1+B|).
\end{equation}
The solution with $z_1>0$ is always present, and it is clear (see Fig. 1),
that in that case (\ref{hesh.2})
is negative. Among the two solutions with $z_1<0$, which exist if
$\beta>\beta_c=n$ and if $h$ is sufficiently small, it is also clear
that for the one with the smaller $|z_1|$ (\ref{hesh.2}) is positive, while
for the one with the larger $|z_1|$ (\ref{hesh.2}) is negative (for the
particular value of $h$ for which these two solutions coincide (\ref{hesh.2})
is zero). Thus, if $n=1$, also one solution with negative $z_1$ is a maximum;
it is the metastable solution, with the magnetization opposite to the
magnetic field, found in the hysteresis curve. We will see below its
metastability, i.e., that for such $z_1$ the exponent in (\ref{MF.Z3}) is
a local maximum, smaller than the absolute maximum obtained for the
positive solution. Now we show that when $n>1$ the only maximum is that
with the positive $z_1$. In fact, for $\mu\neq 1$ we have, for $z_1$
satisfying (\ref{mfsol4}):
\begin{equation}\label{hesh.3}
\left.\frac{\partial^2 f}{\partial z_\mu^2}\right|_{z_\mu=z_1\delta_{\mu 1}}
=-\frac{1}{\beta}\left(1-\frac{z_1}{z_1+B}\right) \quad \quad \mu\ne 1.
\end{equation}
These terms are negative for the solution with positive $z_1$, while they
are positive for the solutions with negative $z_1$ (for which $z_1<z_1+B<0$).

At the end of this computation we can say that in the thermodynamic limit
the specific free energy $F$, obtainable from (\ref{MF.Z3}), is given by
(\ref{freemf}), that is rewritten here:
\begin{eqnarray}\label{freemfa}
-\beta F &=&\lim_{N\rightarrow\infty}\frac{1}{N}\ln Z= \nonumber \\
&=&-\frac{\beta}{2}+
\ln \Omega_{n-1}+\max_{\mathbf{z}}\left[-\frac{1}{2\beta}\mathbf{z}^2 +
\ln G_{n-2}\left(|\mathbf{z}+\mathbf{B}|\right)\right] \nonumber \\
 &=&-\frac{\beta}{2}+
\ln \Omega_{n-1} - \frac{1}{2\beta}\mathbf{z}^{*2} +
\ln G_{n-2}\left(|\mathbf{z}^*+\mathbf{B}|\right).
\end{eqnarray}
The magnetization is simply given, using (\ref{mfsol1}), by
\begin{eqnarray}\label{solmagn}
\mathbf{M}&=&\lim_{N\rightarrow\infty}\frac{1}{N\beta}\frac{\partial}
{\partial \mathbf{h}}\ln Z=\lim_{N\rightarrow\infty}
\frac{\partial}{\partial \mathbf{B}}\left(-\beta F\right)= \nonumber \\
&=&\sum_\mu \left[-\frac{1}{\beta}z^*_\mu \frac{\partial z^*_\mu}
{\partial \mathbf{B}} + g_{n-2}
(|\mathbf{z}^*+\mathbf{B}|)\frac{z^*_\mu+B_\mu}{|\mathbf{z}^*+\mathbf{B}|}
\left(\frac{\partial z^*_\mu}{\partial \mathbf{B}}+\mathbf{\hat{x}_\mu}\right)
\right]= \nonumber \\
&=&\sum_\mu \left[ g_{n-2}
(|\mathbf{z}^*+\mathbf{B}|)\frac{z^*_\mu+B_\mu}{|\mathbf{z}^*+\mathbf{B}|}
\mathbf{\hat{x}_\mu}\right] = \frac{\mathbf{z}^*}{\beta},
\end{eqnarray}
where $\frac{\partial B_\mu}{\partial \mathbf{B}}=\mathbf{\hat{x}_\mu}$ is the
unit vector in the direction of the $\mu$-th axis.
As explained in the main text, while the degeneracy of $\mathbf{z}^*$ when
$h=0$ and $\beta>\beta_c$ does not affect (\ref{freemfa}), the actual
direction of $\mathbf{M}$ in a real system in this case is determined by the
boundary conditions. The equation of state is obtained computing the specific
energy, using (\ref{mfsol1}) and (\ref{solmagn}):
\begin{eqnarray}\label{soleqst}
U&=&-\lim_{N\rightarrow\infty}\frac{1}{N}\frac{\partial}
{\partial \beta}\ln Z= \nonumber \\
&=&\frac{1}{2}-\frac{\mathbf{z}^{*2}}{2\beta^2}+\frac{\mathbf{z}^*}{\beta}\cdot
\frac{\partial \mathbf{z}^*}{\partial \beta}-g_{n-2}\left(|\mathbf{z}^*+
\mathbf{B}|\right)\frac{\mathbf{z}^*+\mathbf{B}}{|\mathbf{z}^*+\mathbf{B}|}
\cdot \left(\frac{\partial \mathbf{z}^*}{\partial \beta}+\mathbf{h}\right)=
\nonumber \\&=&
\frac{1}{2}-\frac{\beta^2\mathbf{M}^2}{2\beta^2}-\frac{\mathbf{z}^*}{\beta}
\cdot \mathbf{h}=\frac{1}{2}(1-M^2)-hM,
\end{eqnarray}
where $M=|\mathbf{M}|$.

We end this Appendix by showing that for $n=1$ the maximum of the exponent
in (\ref{MF.Z3}) with the magnetization opposite to the magnetic field is only
local. We do this with some relations that are valid for any $n$, since they
are useful for the study of the general model. Using (\ref{mfsol1}), the
exponent in (\ref{MF.Z3}) in any stationary point, when this is in the same
direction of the magnetic field (assumed to be the positive direction of the
first axis), is given by:
\begin{equation}\label{maxrel.1}
-\frac{1}{2}g_{n-2}(|x|)\cdot|x|+\frac{1}{2}g_{n-2}(|x|)
\frac{xB}{|x|}+\ln G_{n-2}(|x|),
\end{equation}
where $x=z_1+B$; it can also be written as:
\begin{equation}\label{maxrel.2}
-\frac{1}{2}g_{n-2}(|x|)\cdot|x|+\frac{1}{2}g_{n-2}(|x|)Bsign(x)
+\ln G_{n-2}(|x|).
\end{equation}
The derivative of this expression with respect to $|x|$ gives:
\begin{equation}\label{maxrel.3}
\frac{1}{2}|x|\left(\frac{g_{n-2}(|x|)}{|x|}-g'_{n-2}(|x|)\right)
+\frac{1}{2}g'_{n-2}(|x|)Bsign(x).
\end{equation}
According to the properties of the function $g_n(x)$, the first term
and the coefficient of $sign(x)$ in the second term are positive. Knowing also 
that the solution with magnetization opposite to the magnetic field has a
value of $|x|$ smaller than that of the solution with $h$ and $M$ in the same
direction, this is sufficient to prove that the former, for $n=1$, is only a
local maximum of the exponent in (\ref{MF.Z3}).

\section*{Acknowledgments}
We thank Stefano Ruffo and David Mukamel for the illuminating discussions we
had on some aspects of this work. One of us (D. M.) would like to thank Claudio
Castellani for some suggestions on the matter of subsection \ref{intermat}.

\newpage
\begin{figure}[ht]
\begin{center}
\includegraphics[bb=176 324 427 688]
{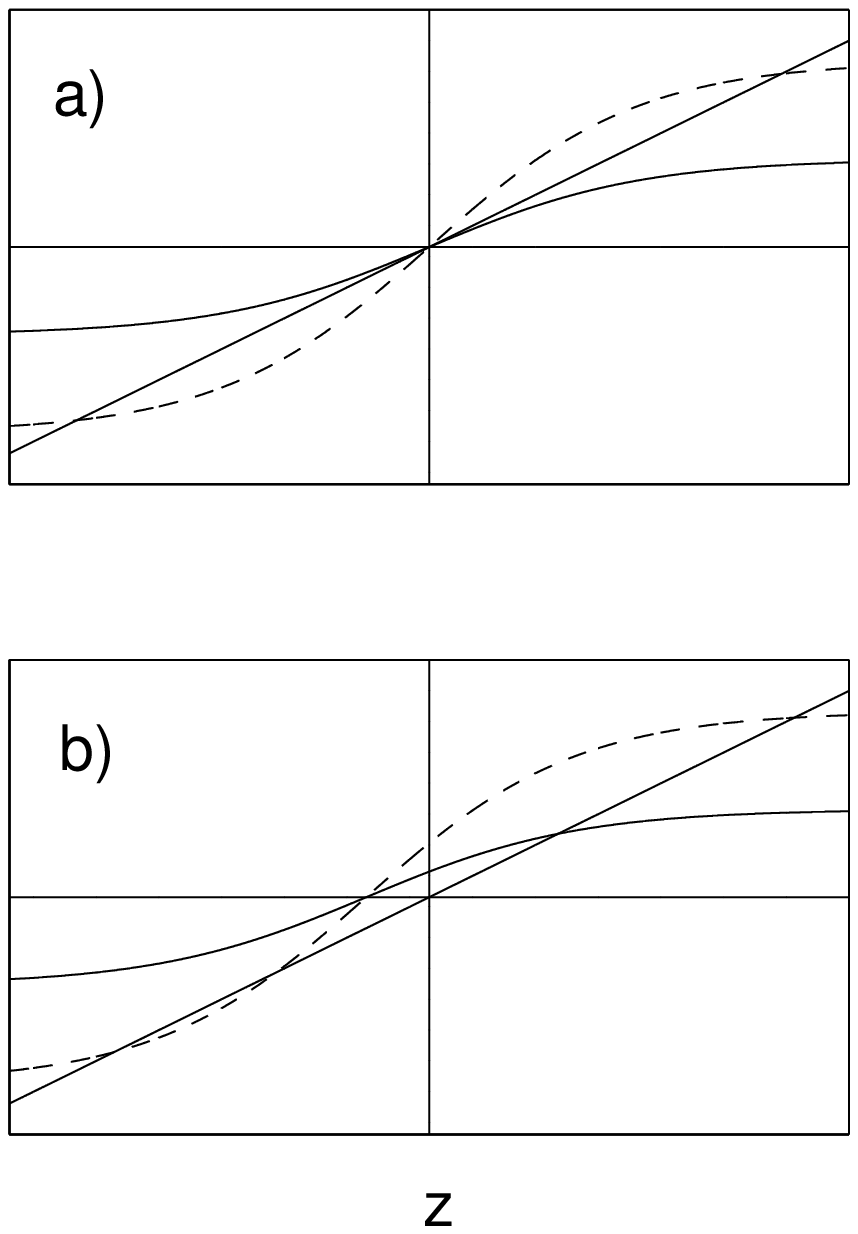}\\
\end{center}
\caption{In both panels the curved lines represent the qualitative behaviour,
for any $n$ and for a range of $z$ about $0$, of $\beta g_n(z+B)$ for
$\beta<\beta_c$ (solid line) and for $\beta>\beta_c$ (dashed line). The
straight line is the bisectrix. a) $B=0$: when $\beta>\beta_c$ the relevant
solution for the modulus $z$ in (\ref{mfsol6}) is the positive intersection, and
not $z=0$; b) $B>0$: the relevant solution for $z_1$ in (\ref{mfsol4}) is
always the intersection with $z_1>0$, even when $\beta>\beta_c$.}
\end{figure}


\begin{thebibliography}{00}
\bibitem{ruehel}
D. Ruelle, Classical statistical mechanics of a system of particles, {\it
Helvetica Physica Acta } {\bf 36}:183-197 (1963).
\bibitem{ruelle}
D. Ruelle, {\it Statistical Mechanics: Rigorous Results} (Imperial College
Press and World Scientific, Singapore, 1999); chap. 3.
\bibitem{gallavotti}
G. Gallavotti, {\it Statistical Mechanics - Short Treatise} (Springer,
Berlin, 1999); chap. 4.
\bibitem{note1}
To be precise, the definitions that we give are those found in the book of
Gallavotti.$^{(3)}$ In Ruelle$^{(2)}$ the definition of stability is the same,
but the definition of temperedness requires
a sufficiently fast decay of only the positive part of the potential energy,
while the negative part is left free. However, the union of the two conditions
is equivalent in both cases.
\bibitem{note2}
The distinction between \emph{extensivity} and \emph{additivity}
has been introduced in a recent paper discussing the
inequivalence of ensembles in the infinite range Blume-Emery-Griffiths
model (see next reference). Before the two terms have been used as synonyms.
Pairwise additive potentials decaying at large distances as
$1/r^{\alpha}$, with $\alpha$
smaller than the embedding space dimension $d$, are called also \emph{
non integrable}.
\bibitem{bmr}
J. Barr\`e, D. Mukamel, and S. Ruffo, Inequivalence of ensembles in a system
with long range interactions, {\it Phys. Rev. Lett.} {\bf 87}:030601 (2001).
\bibitem{ising}
E. Ising, Beitrag zur theorie des ferromagnetismus, {\it Zeitschrift f\"{u}r
Physik} {\bf 31}:253-258 (1925).
\bibitem{kac}
M. Kac, G. E. Uhlenbeck, and P. C. Hemmer, On the van der Waals theory of the
vapor-liquid equilibrium. I. Discussion of a one-dimensional model, {\it J.
Math. Phys.} {\bf 4}:216-228 (1963).
\bibitem{stanley1}
H. E. Stanley, Dependence of critical properties on dimensionality of spins,
{\it Phys. Rev. Lett.} {\bf 20}:589-592 (1968).
\bibitem{cgm}
A. Campa, A. Giansanti, and D. Moroni, Canonical solution of a system of
long-range interacting rotators on a lattice, {\it Phys. Rev. E} {\bf 62}:
303-306 (2000).
\bibitem{fisher64}
M. E. Fisher, Magnetism in one-dimensional systems - The Heisenberg model
for infinite spin, {\it Am. J. Phys.} {\bf 32}:343-346 (1964).
\bibitem{kacberlin}
T. H. Berlin and M. Kac, The spherical model of a ferromagnet, {\it Phys.
Rev.} {\bf 86}:821-835 (1952).
\bibitem{stanley2}
H. E. Stanley, Spherical model as the limit of infinite spin dimensionality,
{\it Phys. Rev.} {\bf 176}:718-722 (1968).
\bibitem{note3}
The basic references are: R. Brout, Statistical mechanical theory of
ferromagnetism. High density behavior, {\it Phys. Rev.} {\bf118}:1009-1019
(1960); M. Kac and E. Helfand, Study of several lattice systems with
long-range forces, {\it J. Math. Phys.} {\bf 4}:1078-1088 (1963);
A. J. F. Siegert and D. J. Vezzetti, On the Ising model with long-range
interaction, {\it J. Math. Phys.} {\bf 9}:2173-2193 (1968).
\bibitem{note4}
This distinction between long and short range interactions
is at variance with the terminology generally used in critical phenomena;
the two differ in the range $d < \alpha < d+2$. See for example: M. E. Fisher
and V. A. Privman, First-order transitions in spherical models: finite-size
scaling, {\it Commun. Math. Phys.} {\bf 103}:527-548 (1986).
\bibitem{tamant}
F. Tamarit and C. Anteneodo, Rotators with long-range interactions: connection
with the mean-field approximation, {\it Phys. Rev. Lett.} {\bf 84}:208-211
(2000).
\bibitem{latrap}
V. Latora, A. Rapisarda, and C. Tsallis, Non-Gaussian equilibrium in a
long-range Hamiltonian system, {\it Phys. Rev. E} {\bf 64}:056134 (2001).
\bibitem{antruf2}
M. Antoni, H. Hinrichsen, and S. Ruffo, On the microcanonical solution of a
system of fully coupled particles, {\it Chaos, Solitons and Fractals}
{\bf 13}:393-399 (2002).
\bibitem{cgm2}
A. Campa, A. Giansanti, and D. Moroni, Metastable states in a class of
long-range Hamiltonian systems, {\it Physica} {\bf A 305}:137-143 (2002).
\bibitem{antruf}
M. Antoni and S. Ruffo, Clustering and relaxation in Hamiltonian
long-range dynamics, {\it Phys. Rev. E} {\bf 52}:2361-2374 (1995).



\end{thebibliography}
\end{document}